\documentclass[twocolumn,floatfix,prx,aps,a4paper,showpacs]{revtex4-2}
\pdfoutput=1

\usepackage{graphicx}
\usepackage{amssymb}
\usepackage{amsmath}
\usepackage[utf8]{inputenc}
\usepackage{cancel}
\usepackage[usenames,dvipsnames]{color}
\usepackage{psfrag}
\usepackage{epsfig}
\usepackage{bbm}
\usepackage{bm}
\usepackage{dsfont}
\usepackage{soul}
\usepackage{colonequals}
\usepackage{float}
\usepackage{enumitem}
\definecolor{TURed}{RGB}{153,0,0}
\definecolor{DarkGreen}{RGB}{0,153,0}
\definecolor{DarkBlue}{RGB}{0,0,153}
\newcommand{\blue}[1]{}
\newcommand{\green}[1]{\textcolor{blue}{#1}}
\newcommand{\nn}{\nonumber\\}

\begin{document}

\title{
Non-reciprocal hidden degrees of freedom:
\\
A unifying perspective on memory, feedback, and activity 
} 
\author{Sarah A.~M.~Loos, Simon~M.~Hermann, and Sabine H.~L.~Klapp}
\affiliation{
  Institut f\"ur Theoretische Physik,
  Hardenbergstr.~36,
  Technische Universit\"at Berlin,
  D-10623 Berlin,
  Germany}
\date{\today}

\begin{abstract}
We show that memory, feedback, and activity are all describable by the same unifying concept, that is non-reciprocal (NR) coupling. 
We demonstrate that characteristic thermodynamic features of these intrinsically nonequilibrium systems are reproduced by low-dimensional Markovian networks with NR coupling, which we establish as minimal models for such complex systems. 
NR coupling alone implies a violation of the fluctuation-dissipation relation, which is inevitably connected to entropy production, i.e., irreversibility. 
Hiding the NR coupled degrees of freedom renders non-Markovian, one-variable Langevin descriptions with complex types of memory, for which we find a generalized second law involving information flow. We demonstrate that non-monotonic memory is inextricably linked to NR coupling. Furthermore, we discuss discrete time delay as the infinite-dimensional limit, %
and find a divergent entropy production, corresponding to unbounded cost for precisely storing a Brownian trajectory.
 \end{abstract}

\maketitle

\section{Introduction}\label{sec:INTRO}
Most classical systems consisting of mutually coupled elements, such as atoms in a polymer, colloids in a dense suspension, or billiard balls,
involve {\em reciprocal} interaction forces which are derivable from a Hamiltonian (i.e., conservative) and thus fulfill, automatically, Newton's third law, {\em actio equal reactio}.
In the absence of driving forces or temperature gradients, such systems can equilibrate and are well described by 
traditional thermodynamics. This holds even on the mesoscale, that is, when instead of the full ``network" of reciprocally coupled elements, only one representative (stochastic) variable, say $X_0$, is
considered by integrating out the other degrees of freedom (d.o.f.), $X_{j>0}$. This is the key idea of the celebrated Mori-Zwanzig approach~\cite{Zwanzig1973} yielding a generalized Langevin equation, which involves a memory kernel satisfying a fluctuation-dissipation relation (FDR) (e.g., for viscoelastic fluids).

However, the idea of reciprocal couplings and its thermodynamic implications breaks down in many living and artificial ``active" systems 
where due to high complexity more general interactions, in particular, symmetry-broken NR couplings between subsystems, or mesoscopic degrees of freedom (d.o.f.), can emerge~\cite{agudo2019active,Durve2018}.
The main goal of this paper is to identify the thermodynamic implications of NR interactions.
Sensing is a typical example for such a NR non-conservative type of interaction: 
The state of one component (the ``sensor''), say $X_1$, depends on the state of the other component (``the measured system''), say $X_0$, but not \textit{vice versa}. Another prominent example is active self-propelled motion due to an (independent) energy input~\cite{ramaswamy2010mechanics,ramaswamy2017active,fodor2016far,marchetti2013hydrodynamics}. A well-know realization are Janus particles with unihemispherical coating~\cite{speck2019thermodynamic}. 
In terms of equations of motion, the temporal evolution of the particle's position 
is assumed to be affected by the noisy orientation, %
but there is no backcoupling. 
Yet another example is \textit{feedback} where the 
elements' response to each other is intrinsically asymmetric, associated with a nonzero 
information flow~\cite{Esposito2016}.
Feedback is an important control mechanism in artificial systems, implemented for example in robots~\cite{Mijalkov2016}, autonomously driving cars~\cite{paden2016survey}, 
or quantum devices~\cite{strasberg2013thermodynamics,Carmele2013}.
A paradigmatic example that can be realized experimentally is delayed feedback control on colloids via optical tweezers setups (see~\cite{Khadka2018,debiossac2019thermodynamics} for experiments and~\cite{Loos2017,Loos2019,Loos2019b,Munakata2014,Rosinberg2015,Rosinberg2017,Longtin2010,van2019uncertainty} for theoretical descriptions).
Feedback loops are also of crucial importance in many biological contexts~\cite{yi2000robust,aguilar2018critical}, e.g., in the metabolism of cells~\cite{schiering2017feedback,pardee1959mechanisms}, neural networks~\cite{Scholl2009}, the chemotaxis of bacteria~\cite{yi2000robust,micali2016bacterial}, and in gene regulatory networks~\cite{lai2016understanding,Josic2011,Rateitschak2007,Friedman2006}.

A common feature of active and feedback systems is that they are intrinsically out of equilibrium already on the level of individual constituents. %
Therefore, both types of systems have been a prime focus of research in nonequilibrium statistical physics. Current attention is particularly concerned with a description via stochastic thermodynamics~\cite{Seifert2012, Sekimoto2010} owing to the noisy character of the individual trajectories. Here, key quantities are the entropy production (EP) and fluctuation theorems, i.e., generalizations of the second law. 
In fact, there is another --less obvious-- shared feature of active and feedback systems, that is, the presence of memory. While the discussion of memory, or \textit{delay}, in feedback loops has a long tradition~\cite{Loos2017,Loos2019,Loos2019b,Munakata2014,Rosinberg2015,Rosinberg2017,Longtin2010,van2019uncertainty} (where it stems e.g. from finite signal processing speed), this viewpoint is less common in the context of active motion. However, the presence of history-dependence becomes apparent in the form of \textit{persistence} of active swimmers~\cite{narinder2018memory,nagai2015collective,Scholz2018}, 
the viscoelastic properties of active gels~\cite{marchetti2013hydrodynamics}, 
or the giant Kovacs-like memory effect in the Vicsek model~\cite{kursten2017giant}.
Despite these similarities, %
the thermodynamics of active~\cite{shankar2018hidden,pietzonka2017entropy,argun2016non,marconi2017heat} and feedback systems~\cite{Loos2019,Munakata2014,Rosinberg2015, Rosinberg2017,van2019uncertainty} %
has essentially been worked out separately. 
Only recently researchers have started to realize the delicate connections between both~\cite{Khadka2018,aguilar2018critical,Mijalkov2016}. We believe that a further cross-fertilization between the fields would be very important not only for the thermodynamic understanding of individual elements, but also for the collective behavior on larger scales; a topic that is already of major interest for active systems~\cite{nagai2015collective,Mijalkov2016}.

In this spirit, we here propose a consistent, unifying thermodynamic description of systems with feedback, activity and memory by modeling them as 
Markovian networks of elements $X_0$, $X_{j>0}$ ($j=1,\dots, n$) with NR coupling and (white) noise. Equivalently, by projecting out $X_{j>0}$,
the dynamics can be formulated as a one-variable system for $X_0$ with a memory kernel and colored noise. 
The main purpose of this paper is to investigate the connection between the topology of the coupling matrix and the thermodynamic properties of the model. To this end, we calculate analytically important thermodynamic quantities such as the energy input into the system, entropy production (EP) and information flows. We find that, ``activity" and ``feedback" can be related to certain properties
of the topology matrix, which are apparent on both levels of description. 
 Conventionally, active systems are mostly associated with ``energy consumption'' of individual constituents~\cite{ramaswamy2010mechanics,ramaswamy2017active,nardini2017entropy}, 
 while feedback-controlled systems are considered to be ``information fueled''~\cite{cao2009thermodynamics}. 
However, also active systems involve nontrivial information flows~\cite{dabelow2019irreversibility,micali2016bacterial}, and a feedback controller clearly ``consumes energy'' in order to operate.
Our approach naturally unifies both perspectives. A key ingredient is the NR coupling yielding nonequilibrium, as expressed by a broken FDR. 

For the sake of generality,
we deliberately do not focus on a specific model, and rather offer different interpretations for the involved d.o.f. 
However, in general, a natural interpretation is that $X_0$ represents the actual d.o.f. of interest, such as the position of a colloidal particle, whereas the other variables $X_{j>0}$ represent those parts of the complex environment which generate the feedback loop or active motion. For a bacterium or active microswimmer with position $X_0$, the other d.o.f. might represent the flagella or asymmetric flow fields~\cite{cates2012diffusive, speck2019thermodynamic}. For a feedback controlled colloid, the $X_{j>0}$ may represent the memory cells of its memory device.
In our approach, we purposefully include these subsystems. 
This conforms with the idea that in real world systems, non-Markovianity in the vast majority of cases stems from coarse-graining \textit{real physical} subsystems. Motivated by their physical nature, we also assume that all subsystems are subject to fluctuations (i.e., coupled to a heat bath), and are governed by the first law of thermodynamics: energy is a conserved quantity, thus, all energy flows associated with each subsystem $X_{j>0}$ must balance out. 
In the case of a feedback controller, this viewpoint contrasts earlier studies, where the memory of a feedback-controller is realized with a \textit{tape}~\cite{parrondo2015thermodynamics,horowitz2014second,horowitz2014thermodynamics,sartori2014thermodynamic,mandal2013maxwell,mandal2012work,barato2013autonomous}\blue{(Maxwell demon models~\cite{mandal2013maxwell,mandal2012work})}. In the interpretation presented here, the heat baths of $X_{j>0}$ account for measurement errors and limited accuracy of the controller's memory device. 
We stress, however, that even in cases where the interpretation of the $X_{j>0}$ may be difficult, various (thermodynamic) properties of $X_0$ are actually {\em independent}
of whether one chooses the non-Markovian description, or the full networks. 
An example is the heat flow. 

From the viewpoint of $X_0$ being the only observable d.o.f., the very idea of representing a time-nonlocal equation by a set of Markovian equations for auxiliary variables $X_{j>0}$
is, of course, not new, and is typically referred to as Markovian embedding~\cite{Siegle2010,Siegle2011,Bao2005,Villamaina2009}. In this regard, the contribution of this paper is to discuss the role of the ``auxiliary'' variables in the thermodynamic description. An obvious advantage is methodological, 
it allows us to derive a large amount of thermodynamic quantities which are notoriously difficult to access
when considering a non-Markovian representation alone such as the total EP, the key measure of irreversibility, and fluctuation theorems. For example, to calculate the total EP one needs the path probability of the time-reversed process, which is {\em acausal} for the (non-Markovian) process
of interest $X_0$ \cite{Munakata2014,Rosinberg2015,Rosinberg2017}. This is because, due to the memory, the reversed process depends on its own --unforseeable-- future.
Recent studies (for systems with exponential, or delta-correlated memory), explore strategies to nevertheless define reasonable irreversibility measures based on the trajectories of $X_0$ only~\cite{caprini2019entropy,dabelow2019irreversibility,Munakata2014,Rosinberg2015,Rosinberg2017,shankar2018hidden}.
While these approaches are promising, a generalization to other types of memory is not straightforward. 
In contrast, our network representation with NR couplings allows for a large number
of complex memory kernels. 
In all cases, the Markovianity of the full network implies the existence of a 
closed Fokker-Planck equation, and standard expressions for path probabilities, allowing to access the total EP via well-understood formalisms~\cite{Seifert2012, Sekimoto2010}.  This new perspective on Markovian embeddings, and, in particular, on the connection between topology matrix and the memory, is a further important contribution of this paper.

Since control has been a central topic in thermodynamics for many years~\cite{parrondo2015thermodynamics}, we add some remarks differentiating our approach from earlier studies. First, we stress that we treat continuous, stochastic systems, as opposed to earlier bipartite-models basing on Master-equations like~\cite{hartich2014stochastic,boukobza2006thermodynamics,horowitz2014thermodynamics}. We further consider autonomous systems~\cite{horowitz2014thermodynamics}, described by
\textit{time-continuous} delayed equations. This contrasts ``Maxwell-demon'' protocols as in~\cite{Sagawa2010,sagawa2012nonequilibrium,cao2009thermodynamics,Esposito2016}, where the controller only interacts with the system $X_0$ at discrete instances in time, such that in between, the dynamics of $X_0$ is Markovian and has a causal time-reversed process. To emphasize this difference, the type of control considered here is also denoted \textit{non-Markovian feedback control}~\cite{debiossac2019thermodynamics}. Important theoretical contributions to this field were done by Rosinberg and Munakata~\cite{Munakata2014,Rosinberg2015,Rosinberg2017}, %
who lately introduced a framework to define irreversibility measures (based on $X_0$ alone), for the case of underdamped motion with \textit{discrete }delay. 
This is indeed a particularly difficult case, due to the infinite-dimensional nature of the memory kernel (a delta-distribution).
Our approach includes, in fact, discrete delay as a limiting case, putting it in a different perspective. 
Further, the case of delayed feedback control with \textit{measurement errors} was rather rarely considered so far~\cite{Munakata2013,horowitz2014second,Argun2016}.
We close the introduction with a brief outline of the remainder of the paper which, at the same time, gives an overview of the main results.
After introducing the minimal models in Sec.~II, we show in Sec.~III that NR coupling alone yields non-equilibrium. To this end, we discuss the (broken) FDR, and show that the system displays a positive EP. By considering the energy flows through the system in Sec.~IV, we then establish a connection between NR coupling and activity. In Sec.~V, we analyze the system from an information-theoretical perspective. We identify information flows accompanying the energy flows, which allow for a clear definition of sensing, active propulsion and feedback. We further derive a generalized second law for active and feedback-controlled systems.
In Sec.~VI. we discuss the thermodynamics in the presence of non-monotonic memory, in particular, the heat flow and total EP as functions of the network size. We also consider the EP fluctuations which fulfill an integral fluctuation theorem, and discuss discrete delay as a limiting case.

%
%
%
%
\section{Minimal models}
We consider Markovian Langevin networks of the type
\begin{subequations}\label{eq:Network}
\begin{align}
\gamma_0\dot{X}_0 &= a_{00} X_0 +a_{0n} X_n + f_0 + \xi_0, \\
\gamma_j\dot{X}_j &= a_{jj} X_j +a_{j-1 j} X_{j-1} +  \xi_j,  \label{eq:Network_Chain}
\end{align}
\end{subequations}
with $ j\in \{1,...,n\},$ Gaussian white noises $\langle \xi_i(t)\xi_j(t') \rangle =2 k_\mathrm{B}\mathcal{T}_j \gamma_j \,\delta_{ij}\delta(t-t')$ at temperatures $\mathcal{T}_j$, and $k_\mathrm{B}$, $\gamma_j$ being the Boltzman and friction constants. $f_0$ is an, in general, nonlinear force.
The topology matrix $\underline{\underline{a}}$ defines the strength of the couplings $a_{ij}$, and gives the timescale $\gamma_j/a_{jj}$ of the {exponential relaxation dynamics} of each d.o.f., due to the restoring forces $a_{jj}X_j$. (We will in Sec.~\ref{sec:non-Monotonic}, in fact, also briefly discuss more general coupling schemes, where $X_j$ is additionally coupled to $X_{j+1}$.)

Hiding the $X_{j>0}$, i.e., projecting them onto $X_0$\blue{via Mori-Zwanzig projection method}
 as described in~\cite{Loos2019b,Zwanzig1973}, yields 
the overdamped \textit{non-Markovian} Langevin equation (LE)
\begin{align} \label{eq:LE-X0}
\gamma_0\dot{X}_0(t) =&a_{00}X_0(t) + a_{0n}\int_{0}^{t} K(t-t') X_0(t')\mathrm{d}t'\nn 
&+ f_0+ \nu(t)  + \xi_0(t) ,
\end{align}
where the second term involves a time-nonlocal force
depending on the \textit{past} trajectory, weighted with a memory kernel $K$, and scaled with the strength $a_{0n}$. We aim to emphasize that the dynamics of $X_0$ is \textit{identical} to~(\ref{eq:Network}). Using (\ref{eq:LE-X0}) as opposed to (\ref{eq:Network}) can be regarded as a coarse-graining, because the dynamics of $X_j$ is not explicitly considered. However, it does \textit{not} imply loss of information about (or approximation of) $X_0$. Since we are interested in analytical solutions, we will focus  in this paper on the linear case, i.e., $f_0=0$, but the framework is readily adaptable to nonlinear cases.

The details of the memory kernel and the correlations of the zero-mean, Gaussian colored noise $\nu$ depend on the topology of the coupling matrix (as discussed below). For $\mathcal{T}_{j>0}\equiv 0$, there is no colored noise in~(\ref{eq:LE-X0}).
In the following, we show that a \textit{non-reciprocal (NR) coupling} $a_{ij} \neq a_{ji}$, in~(\ref{eq:Network}), and, in particular \textit{unidirectional coupling} $a_{ij}\neq 0, a_{ji}=0$, has important implications for the thermodynamical and dynamical properties of $X_0$, including the possibility to describe more complex types of memory and model feedback control, as well as \textit{active} systems.

Concerning the meaning of the NR interactions, we aim to emphasize the following. Fundamentally, physical interactions are typically reciprocal and symmetric, including mechanical coupling, as states Newton's third law. But, 
in complex systems, involving \textit{chemical reactions}, multiscale processes, or coupling between a chemical component and a mechanical d.o.f., much more complicated relationships between subsystems can emerge~\cite{agudo2019active,Durve2018,Kompaneets2008,Ivlev2015} (which might be representable as coupling between few mesoscopic, stochastic d.o.f.). This is the situation considered here. 

We deliberately do not specify the physical nature of the individual components, but keep them abstract. The reason for that is twofold. First, we aim at understanding the implications of NR coupling for the thermodynamic and dynamic properties on a general level. Furthermore, we aim to unify different perspectives and point out a deep connections between activity, feedback, memory and control; rather than analyze a specific system. 
However, to support our arguments and illustrate the idea, we give some examples for possible interpretations of the $X_j$.
%
%
%
\subsection{Examples}\label{sec:examples}
We begin by considering the impact of a \textit{single} NR interaction. In particular, we study 
\begin{itemize}
\item[(I)] 
the smallest version of (\ref{eq:Network}) with $n=1$, where the memory kernel and the noise correlations $C_\nu(T) := \langle \nu(t)  \nu(t+T) \rangle$  decay exponentially,
\begin{align} \label{eq:ExampleI}
C_\nu(T)  &=  k_\mathrm{B}\mathcal{T}_1 (a_{01}^2/a_{11}) e^{a_{11}T/\gamma_1},\nn
K(T) &=   (a_{10} /\gamma_{1}) e^{a_{11}T/\gamma_1}.\tag{I}
\end{align}
\end{itemize}
For unidirectional coupling ($a_{10}=0$, $a_{01}> 0$), (\ref{eq:ExampleI})  
describes the dynamics of a \textit{microswimmer} at position $X_0$ within the active Ornstein-Uhlenbeck particle (AOUP)-model~\cite{shankar2018hidden,dabelow2019irreversibility,caprini2019entropy,Mandal2017,bonilla2019active}. $X_1$ models the effect of the flagella of a bacterium, or the asymmetric flow field around a Janus colloid, responsible for the propulsion force on the swimmer, while the memory models the persistence of the motion. Remarkably, the very same network with reversed unidirectional coupling (i.e., $a_{01} \leftrightarrows a_{10}$), was recently suggested as a model for a \textit{cellular sensor}~\cite{bo2015thermodynamic,hartich2016sensory}.

Thermodynamic properties of a two d.o.f. network with exponential memory were investigated in~\cite{Crisanti2012,Puglisi2009}. We here generalize to arbitrary systems sizes ($n$) and coupling topologies, and change the focus onto NR interactions. Still, we use (\ref{eq:ExampleI}) as an illustrative example, as it is the most simple one. 

Further, we discuss networks with multiple NR interactions. In particular, we consider
\begin{itemize}
\item[(II)] 
a unidirectionally coupled ring of length $n$, coupled to $X_0$ with $a_{0n}=k$, $a_{i i-1}=-a_{ii}=\gamma' n/\tau$, $\mathcal{T}_{j>0}=\mathcal{T}'$, $\gamma_{j>0}=\gamma'$ 
yielding colored noise and Gamma-distributed memory (see the Appendix~\ref{sec:deriveKandNu} for a derivation),
\begin{align}\label{K_noise_II}
C_\nu(T)    &=   k^2  \frac{k_\mathrm{B}\mathcal{T}'}{\gamma'}  \sum_{p=0}^{n-1} \sum_{l=0}^{p} 
 \frac{2^{l-2p}(2p-l)!}{p!(p-l)!l!} \frac{e^{-n T /\tau} T^l}{(\tau/n)^{l-1}} ,
\nn
K(T) &= \frac{n^n \, T ^{n -1}}{\tau^n(n-1)!}  e^{-\frac{nT}{\tau} }. \tag{II}
\end{align} 
\end{itemize}

An important application of (\ref{K_noise_II}) is a colloidal particle $X_0$ under ``optical tweezers''-feedback control with strength $k$, where $X_{j>0}$ model the memory cells of the controller, see Sec.~\ref{sec:non-Monotonic}.
The case $n=1$ is included in example~(\ref{eq:ExampleI}). For $n>1$, the memory kernel has a unique maximum around its mean at $\tau$, and a variance $\tau^2/n$ which decreases with $n$, as plotted in Fig.~\ref{fig:kernels} for $n=1,2,3$. \textit{Discrete} delay, i.e., $K(T)\to\delta(T-\tau)$, is approached in the limit $n\to\infty$~\cite{Loos2019b}.

Furthermore, also memory kernels with multiple extrema can be modeled by linear networks, for example by coupling two rings of type (\ref{K_noise_II}) to $X_0$, see Fig.~\ref{fig:kernels} (d) for an example. We observe that a kernel with $N$ extrema can be represented via (at least) $N$ d.o.f., which is because each d.o.f. possibly introduces one new timescale. On the other hand, we observe that the colored noise produced by linearly coupled d.o.f. is always monotonically decreasing (see Fig.~\ref{fig:kernels}). 
 
In Sec.~\ref{sec:non-Monotonic}, we will discuss the importance of NR interactions for the generated memory in more detail. In the following Sec.~\ref{sec:noneq}, we will first explore the consequences of NR coupling for the thermodynamic properties of the model, focusing on the asymptotic behavior ($t\to\infty$), when the system approaches a {steady state}. 
%

\section{NR coupling $\&$ Non-equilibrium}\label{sec:noneq}
Here, we demonstrate that NR coupling alone ``drives'' the system out of equilibrium. To this end, we consider two typical measures, first, the fluctuation-dissipation relation (FDR) for non-Markovian systems, and the total entropy production (EP).
\subsection{Fluctuation-Dissipation relation}\label{sec:FDR}
The FDR (or Kubo relation of second kind)~\cite{Kubo1966} 
\begin{equation}\label{eq:FDT}
\langle \mu(t)\mu(s)\rangle {=} k_\mathrm{B} \mathcal{T}_0 \, \gamma(|t-s|),
\end{equation}
describes a balance between friction kernel $\gamma$ and thermal noise $\mu$. As well known for e.g., viscoelastic-fluids, the validity of a FDR
implies that the system equilibrates in the absence of external driving~\cite{Kubo1966,Maes2013}. 
To check~(\ref{eq:FDT}) for the present model, we rewrite~(\ref{eq:LE-X0}) in the form of a generalized LE 
by converting $K$ via partial integration into a friction kernel,
obtaining
\begin{align} 
\int\limits_{0}^{t} \gamma(|t-s|) \dot{X}_{0}(s)\,\mathrm{d}s=& a_{00}X_0 + \widetilde{K}(0) {X}_{0}(t)+\mu(t),
\end{align} 
involving $\mu(t)= \xi_0(t) +  \nu (t)$ and 
\begin{align} 
\gamma(t-s)=&  2\gamma_0 \, \delta(t-s) + {\widetilde{K}(t-s)} ,
\end{align} 
where $\widetilde{K}(T)=  (a_{01}a_{10}/a_{11}) e^{a_{11} T/\gamma_1}$ for case (\ref{eq:ExampleI}), and %
$\widetilde{K}(T)=k\,\Gamma\left( n, {n  T}/{\tau} \right)/(n-1)!$ for case (\ref{K_noise_II}), %
with the upper incomplete gamma function. For $n=1$, it can easily be verified that the FDR holds if
\begin{align} \label{eq:Fulfill-FDR}
a_{01} \mathcal{T}_1=a_{10} \mathcal{T}_0 .
\end{align} 
Moreover, it is violated for all cases (\ref{K_noise_II}) with $n>1$ 
(see the Appendix~\ref{sec:FDT_modelII} for a proof). 
Remarkably, (\ref{eq:Fulfill-FDR}) suggests that the system with NR coupling and $n=1$ can reach equilibrium despite $\mathcal{T}_0 \neq \mathcal{T}_1$. This is in sharp contrast to reciprocally coupled (``passive'') systems, which generally never equilibrate in the presence of temperature gradients.
However, this is only true for NR coupling where both couplings are nonzero and have the same sign, i.e., $a_{01}a_{10}>0$. We will later in Sec.~\ref{sec:Energy} show that this case corresponds to a ``mild'' form of symmetry-breaking.
\begin{figure}
\includegraphics[width=0.5\textwidth]{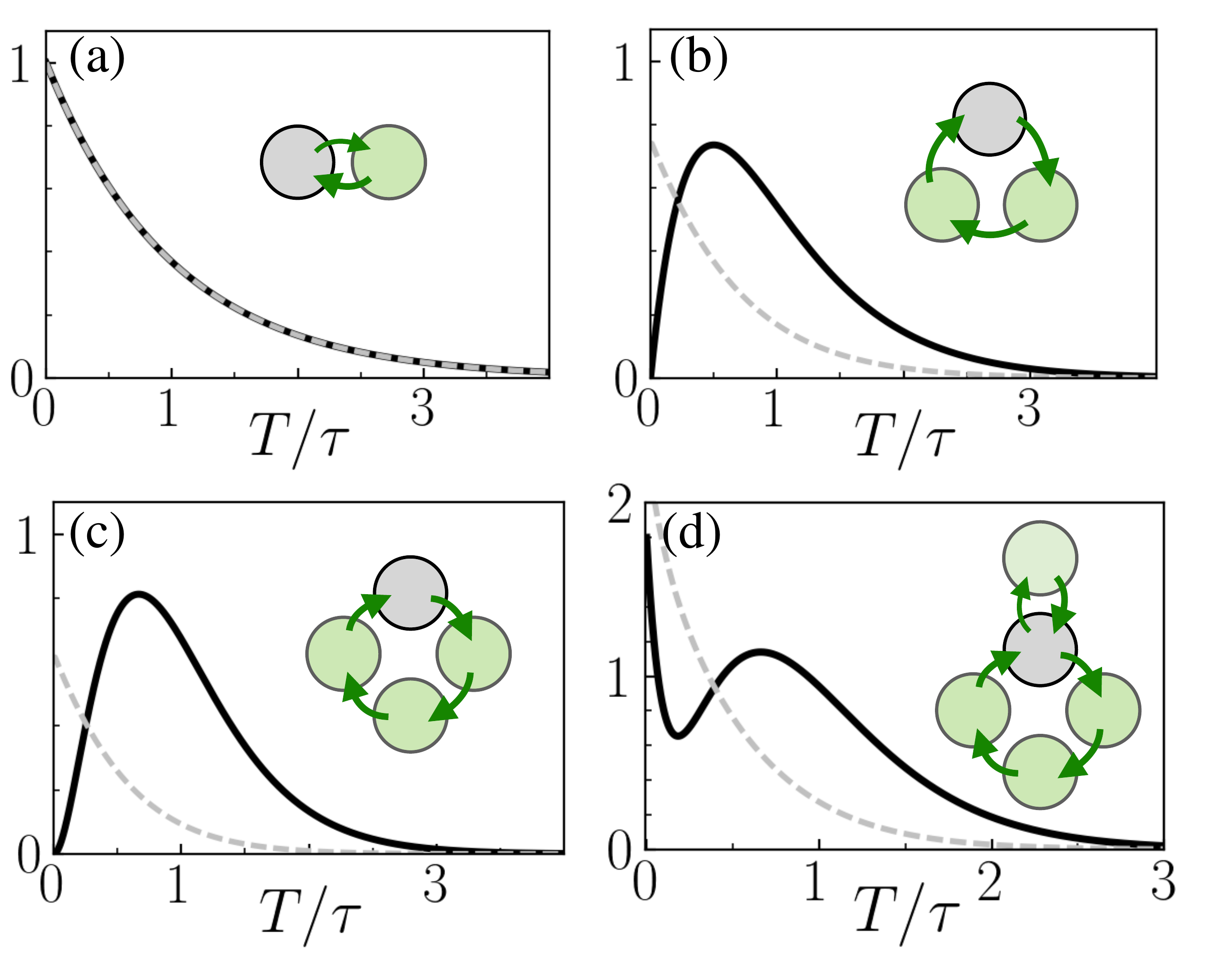} 
\caption{Memory kernels $K(T)$ (black, solid lines) and noise correlations $\nu_n$ (dashed, gray lines) generated by Markovian networks with different topology and size (insets). (a, b, c): networks of type (\ref{K_noise_II}) with $n=1,2,3$, respectively. Other parameters:  $k=-1$, $\gamma=\gamma'=1$. (d): in addition to a linear ring with $n=3$, $k=-4$, a second ring of type (\ref{K_noise_II}) is NR coupled to $X_0$ with $n=1$, $k=-1/2$, $1/10 \tau$. 
} \label{fig:kernels}
\end{figure}
\subsection{Total entropy production}
We now consider the total entropy production (EP), which gives a more fundamental notion of non-equilibrium. While FDR indicates non-equilibrium, it does not yield a reason\textit{ why} the system is out of equilibrium, nor does it quantify the distance from equilibrium. 

However, calculating the EP for a non-Markovian process~(\ref{eq:LE-X0}) is not straightforward~\cite{Munakata2014,Rosinberg2015,Rosinberg2017,Crisanti2012}. In particular, as there is no closed Fokker-Planck equation (FPE) to~(\ref{eq:LE-X0}), and no standard path integral formalism (the Jacobian of the transformation $\xi\to X$ is in general a highly nontrivial path-dependent functional, see~\cite{Munakata2014,Rosinberg2015,Rosinberg2017}), both standard routes~\cite{Seifert2012} are not readily applicable. There are different approaches to the problem, but not all yield consistent results, see~\cite{Crisanti2012} for an example. We will, thus, instead exploit from now on the Markovian version of our model~(\ref{eq:Network}), which allows application of standard framework. 
%
%
%
%
%
%
%
\begin{figure}
\includegraphics[width=0.5\textwidth]{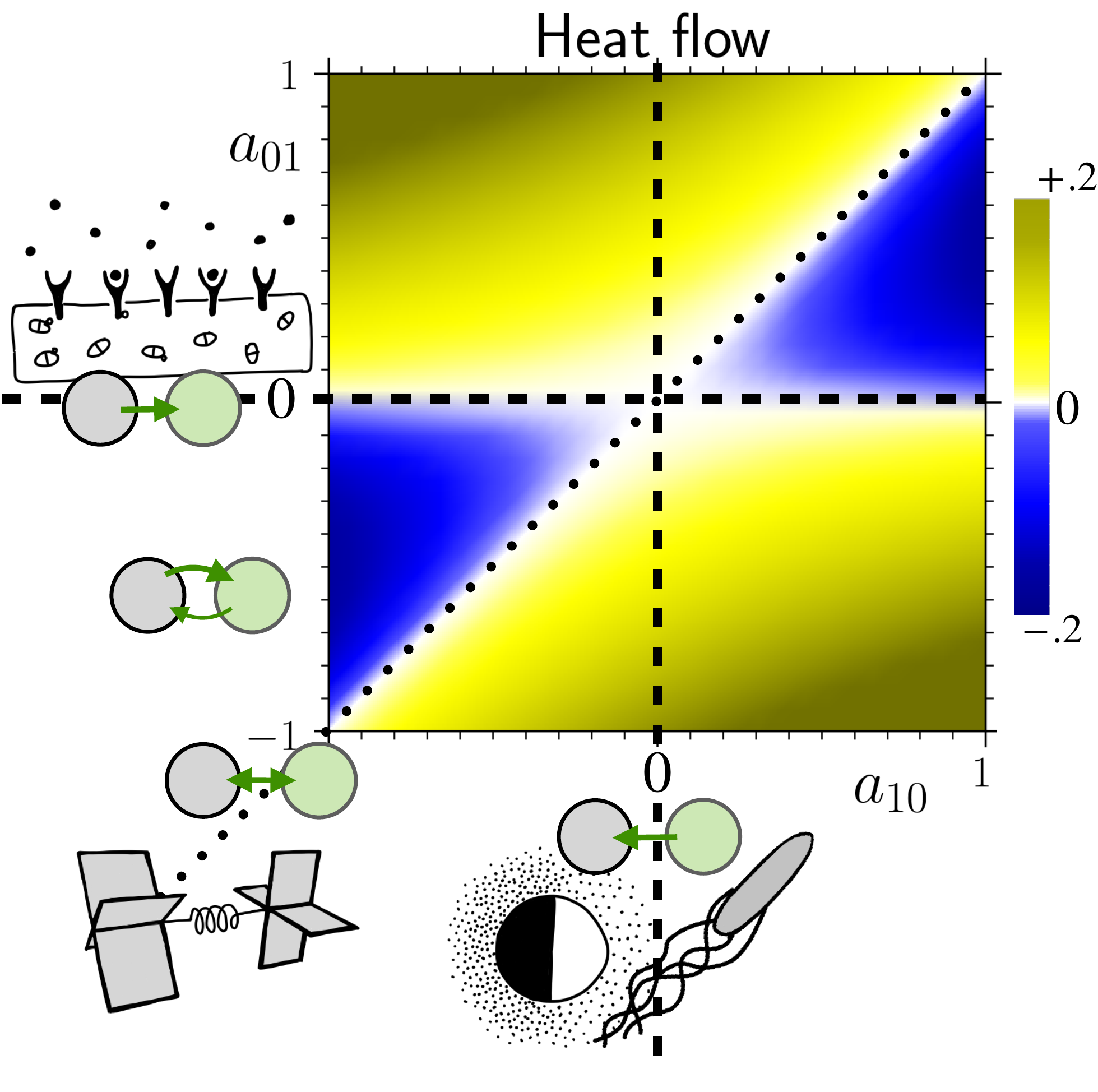}
\caption{Steady-state mean heat flow $\dot{Q}_0$~(\ref{eq:Q0_n=1}) in the minimal model (\ref{eq:ExampleI}) with $a_{11} = a_{00} =-1$, at isothermal conditions $\mathcal{T}_0=\mathcal{T}_1=1$, and $\gamma_0=\gamma_1$. 
$\dot{Q}_0$ vanishes along the diagonal $a_{01}\!=\!a_{10}$, where the system is reciprocally coupled and, thus, ``passive'' (in this case, $X_{0,1}$ could, e.g., model the angles of vanes coupled by a torsion spring~\cite{Sekimoto2010}). $\dot{Q}_0$ further vanishes on the unidirectional coupling line $a_{01}=0$, where $X_0$ is ``sensed'' by $X_1$ (corresponding to a cellular sensor~\cite{hartich2016sensory}). $\dot{Q}_0>0$ along the other unidirectional coupling line $a_{10}=0$ (where $X_0$ corresponds to the position of an active swimmer in the AOUP model~\cite{shankar2018hidden,dabelow2019irreversibility,caprini2019entropy,Mandal2017,bonilla2019active}, $X_1$ modeling the effect of a flagellum of a bacterium, or the asymmetric flow field around a Janus particle). All other parameters are set to unity.}\label{fig:heat}
\end{figure}
%
As well-known, 
the {total EP} along a fluctuating trajectory $\mathbf{\underline{X}}$ $=\{{X}_0(t'),..,X_n(t')\},$ $t'\in [{t_{s}},{t_{f}}]$ is given by~\cite{Lebowitz_1999,Seifert2012}
\begin{align}\label{eq:Stot_general}
\frac{\Delta s_\mathrm{tot}[\mathbf{\underline{X}}]}{k_\mathrm{B}} &  =  \ln \frac{ \mathcal{P}[ \mathbf{\underline{X}} ] }{ \mathcal{\hat{P}}[ \hat{\mathbf{\underline{X}}} ] }
=
\frac{\Delta s_\mathrm{sh}}{k_\mathrm{B}}+ \ln \frac{ \mathcal{P}[ \mathbf{\underline{X}} |\underline{x}_s] }{ \mathcal{\hat{P}}[ \hat{\mathbf{\underline{X}}} |\underline{x}_f] },
\end{align}
involving the {multivariate} (joint) Shannon entropy $s_\mathrm{sh}=-k_\mathrm{B}\ln[ \rho_{n+1}(\underline{x})]$ of the $(n+1)$-point joint probability density function (pdf) $\rho_{n+1}$, and the path probabilities $\mathcal{P}$ and $\mathcal{\hat{P}}$ for forward and backward process. 
The total entropy production~(\ref{eq:Stot_general}) directly quantifies the irreversibility, i.e., time-reversal symmetry breaking. By construction, $\Delta s_\mathrm{tot}$ fulfills an integral fluctuation theorem, $\langle e^{-\Delta s_\mathrm{tot}/k_\mathrm{B}}\rangle =1$, as demonstrated in the Appendix~\ref{sec:IFT}, and confirmed by numerical simulations in Fig.~\ref{fig:Distributions}. This implies that the second law of thermodynamics $\Delta S_\mathrm{tot}\geq 0$, where the capital letter denotes the ensemble average (as in the following).
 $\mathcal{P}$ conditioned on the starting point $\underline{X}(t_{s})=\underline{x}_{s}$, is given by 
the exponential of the Onsager-Machlup action~\cite{Seifert2012, Puglisi2009}. The later readily follow from the Langevin Eqs.~(\ref{eq:Network}), and read
\begin{eqnarray}\label{eq:Path-prob}
\mathcal{P}[ \mathbf{\underline{X}} |\underline{x}_s]\propto  & e^{- \sum_{j=0}^{n} \int_{t_s}^{t_f}   \frac{ \big(\gamma_j \dot{X}_j(t')-F_j[\underline{X}(t')] \big)^2}{4 k_\mathrm{B}\mathcal{T}_j \gamma_j }  \mathrm{d} t' },\nn
\mathcal{\hat{P}}[ \hat{\mathbf{\underline{X}}} |\underline{x}_f] \propto & \,e^{- \sum_{j=0}^{n} \int_{t_s}^{t_f}   \frac{ \big(-\gamma_j \dot{X}_j(t')-F_j[\underline{X}(t')] \big)^2}{4 k_\mathrm{B}\mathcal{T}_j \gamma_j }  \mathrm{d} t' }.\label{EQ:Path-prob-rev}
\end{eqnarray} 
with $F_{j>0}  = a_{jj}X_j +a_{j-1 j} X_{j-1}$, and $F_0=a_{00}X_0 + a_{0n} X_{n}.$
We assume that $X_{j>0}$ are even under time-reversal, like positions, angles, or orientations, which is more suitable for the examples considered in this paper 
(e.g., when $X_{j>0}$ model memory cells of a feedback-controller as we will argue below, odd parity would imply information storage in particle velocities, which seems less appropriate.)
{For active swimmers, the parity of $X_{j>0}$ is indeed a nontrivial aspect, and subject of ongoing debate, see e.\,g.,~\cite{shankar2018hidden}.}
This assumption implies
\begin{equation}\label{eq:Stot-extended}
\ln \frac{ \mathcal{P}[ \mathbf{\underline{X}} |\underline{x}_s] }{ \mathcal{\hat{P}}[ \hat{\mathbf{\underline{X}}} |\underline{x}_f] }= \int_{t_s}^{t_f} \frac{ \delta q_0}{\mathcal{T}_0}+ \sum_{j=1}^{n} \int_{t_s}^{t_f} \frac{ F_j \circ \mathrm{d}X_j}{\mathcal{T}_j} ,
\end{equation}
%
where $\delta q_0$ can be identified as the heat flow from $X_0$ to its bath~\cite{Sekimoto2010}, i.e., the dissipation of $X_0$, which reads 
$\delta q_0   = [a_{00}  X_0 + \,a_{0n} X_n  ]\circ \mathrm{d}X_0$. In the steady state, $\Delta S_\mathrm{sh}=0$, and the ensemble average of the EP~(\ref{eq:Stot_general}) is thus
\begin{align}\label{eq:meanStot_general}
 \dot{S}_\mathrm{tot}   =&
   \frac{a_{0n}^2 \langle  X_n^2 \rangle + a_{0n} \,a_{00} \langle  X_0 X_n  \rangle }{\mathcal{T}_0} 
\nn
&  +
\sum_{j=1}^{n}\frac{a_{j j-1}^2  \langle X_{j-1}^2 \rangle + a_{jj}a_{jj-1}\langle X_{j-1} {X}_{j}  \rangle }{\mathcal{T}_j}
,
\end{align}
where we have used that $X_j \dot{X}_j$-correlations vanish in the steady state, since $  2 \langle X_i\dot{X_i}\rangle= \mathrm{d}\langle X_i^2\rangle/\mathrm{d}t= 0$. The mean EP~(\ref{eq:meanStot_general}) has two contributions, that is, the dissipation of $X_0$
\begin{align}\label{eq:Q0_general}
\dot{Q}_0 = a_{0n}^2 \langle  X_n^2 \rangle + a_{0n} \,a_{00} \langle  X_0 X_n  \rangle , 
\end{align}
and a contribution from the additional d.o.f. $X_{j>0}$.

From the perspective of a non-Markovian process~(\ref{eq:LE-X0}), the second contribution can be considered as the ``entropic cost'' of the memory~\cite{Munakata2014,Rosinberg2015,Rosinberg2017}. At this point it should be emphasized, however, that memory is \textit{not} unequivocally connected to EP. For example, the symmetric case ($a_{01}=a_{10}$) of our network (\ref{eq:ExampleI}) also generates memory in $X_0$, but no net dissipation, thus no EP (see below). 
We recall that there are, of course, real-world examples of physical system which involve complex memory but nevertheless reach thermal equilibrium (where the mean EP naturally vanishes), e.g., viscoelastic fluids. 
As we will later see, such ``entropic cost''-free memory is, however, not useful in terms of extracting net information or for achieving goals of feedback control.

From~(\ref{eq:meanStot_general}, \ref{eq:Q0_general}), closed expressions for $\dot{Q}_0$ and the total EP can be derived for any $n$, as explained in Sec.~\ref{sec:analytical_solutions}. Later, we will discuss the EP and the heat flow for cases $n\geq 1$. Since the resulting expressions are very cumbersome, we report at this point only the results for the case $n=1$ (which was also discussed in~\cite{Crisanti2012}), which read
\begin{align}
 \dot{S}_\mathrm{tot} & =  k_\mathrm{B} \frac{( a_{10}\mathcal{T}_0  - a_{01} \mathcal{T}')^2}{\mathcal{T}_0\mathcal{T}_1(-a_{00}/\gamma_0 -a_{11}/\gamma_1)},\label{eq:Stot_n=1} \\
 \dot{Q}_\mathrm{0} & =\frac{a_{01}(a_{10}\mathcal{T}_0  - a_{10} \mathcal{T}_1)}{(a_{00} +a_{11}\gamma_0/\gamma_1)}. \label{eq:Q0_n=1}
\end{align}
From~(\ref{eq:Stot_n=1}) one immediately sees that the EP vanishes if, and only if, the FDR~(\ref{eq:Fulfill-FDR}) is fulfilled. In this case, also the heat flow~(\ref{eq:Q0_n=1})
and, in fact, the total energy input~(\ref{eq:Einput}) which we discuss below, vanish. Results for $\dot{Q}_0$ are illustrated in Fig.~\ref{fig:heat}, which highlights the role of NR coupling.\blue{[The total EP~(\ref{eq:Stot_n=1}) is plotted in Fig.~\ref{fig:entropy_n=1-2}.]} We conclude that the notion of equilibrium based on the non-Markovian description (from FDR) and the network representation with additional d.o.f. are consistent. This finding is in accordance with~\cite{nardini2017entropy}, but contrasts the viewpoint in~\cite{Crisanti2012}, where a different definition of the total EP was used. We further note an interesting observation concerning the so called ``non-Markovian FDR'' from~\cite{Zwanzig1973}, $\nu(T) = \langle X_0^2\rangle\,K(T)$. This relation is not suitable to correctly describe the equilibrium here, as it is broken despite zero EP. We suspect that this is due to the overdamped limit, which requires velocity relaxation as an {additional} assumption.

While the non-equilibrium nature of the non-Markovian Eq.~(\ref{eq:LE-X0}) is indicated by the violation of FDR~(\ref{eq:Fulfill-FDR}), the mechanism that prevents the system from equilibration, i.e., the driving, is not obvious. As we will show in Sec.~\ref{sec:Energy}, this point is clarified by studying the connection between network topology and energy flows in the corresponding Markovian description.

Since we are primarily interested in the role of NR coupling, we will focus in the remainder of this paper on isothermal conditions, i.e., $\mathcal{T}_{j}=\mathcal{T}$ for all $j>0$. 

\subsection{Analytical solutions \label{sec:analytical_solutions}}
As already indicated by~(\ref{eq:meanStot_general}, \ref{eq:Q0_general}) for the total EP and the heat flow, various (thermo-)dynamic quantities can be calculated on the basis of (cross-)correlations $\langle X_iX_{j} \rangle $, including the pdfs, and information flows [see~(\ref{eq:Info-flow-0})]. For example, the steady-state pdf $\rho_{n+1}$ 
is, due to the linearity of the model, a Gaussian-distribution with zero mean and the covariance matrix $(\underline{\underline{\Sigma}})_{i j}=\langle X_i X_j\rangle$. Thus, it is fully determined by all the (cross-)correlations.
 
Here we thus sketch how closed expressions for the correlations can be obtained for arbitrary network sizes $n$.
To this end, we transform Eqs.~(\ref{eq:Network_Chain}) via the Fourier-transformation
 $\tilde{X}_j(s)=\int_{-\infty}^{\infty} X_j(t)e^{-i \omega t}\mathrm{d}t$, which readily yields %
 \begin{align}\label{eq:Laplace_Xjb}
 i \omega \underline{\underline{\gamma}}  {\underline{\tilde{X}}}(\omega)=& \underline{\underline{a}} {\underline{\tilde{X}}}(\omega) + {\underline{\tilde{\xi}}(\omega) }
 \\
\Rightarrow {\underline{\tilde{X}}}(\omega)=& \underbrace{\left(  i \omega \underline{\underline{\gamma}} -\underline{\underline{a}}\right)^{-1} }_{=\underline{\underline{\tilde{\lambda}}}(\omega)
 } \,{\underline{\tilde{\xi}}(\omega) }.
 \end{align} 
with the Green's function in Fourier-space $\underline{\underline{\tilde{\lambda}}}(\omega)$, determined by the inverse of the topology matrix and the diagonal friction matrix $\underline{\underline{\gamma}}$ with elements $\gamma_j$.
Using a well-known relationship between spatial correlations and the Green's function from linear response theory~\cite{Hanggi1982}, one readily finds
\begin{align}\label{eq:Solution-Corr}
\langle X_j^2\rangle & = \frac{k_\mathrm{B}}{\pi} \sum_{p=0}^{n} \mathcal{T}_p \gamma_p\int_{-\infty}^{\infty} \tilde{\lambda }_{jp}(\omega) \tilde{\lambda }_{jp} (-\omega) \,\mathrm{d}\omega,
\\
\langle X_jX_{l}\rangle & = \frac{k_\mathrm{B}}{\pi} \sum_{p=0}^{n} \mathcal{T}_p \gamma_p \int_{-\infty}^{\infty} \tilde{\lambda }_{jp}(\omega) \tilde{\lambda }_{l p} (-\omega) \,\mathrm{d}\omega.
\end{align}
These are closed expressions for all correlations for arbitrary network sizes. 
The matrix inversion is indeed possible up to very large network sizes due to the sparse coupling. To evaluate the integrals, the residue theorem can be used. However, this requires finding the roots of a polynomial of order $n+1$. Using computer algebra systems, this can be done reasonably fast up to about $n=10$. We also note, for the case $\mathcal{T}_{j>0}=0$, solutions up to $n\sim 10^4$ can be found in this way. 
\section{NR coupling $\&$ activity}\label{sec:Energy}
To further unravel the non-equilibrium nature of the systems with NR coupling, we consider the system from an energetic perspective. 
Since all d.o.f. are of physical nature,
each force applied by, or to, a subsystem $j$ is inevitably connected to an energy exchange. 
Considering the energy associated with a coupling from a mechanical point of view, one finds that there is the fluctuating work~\cite{Sekimoto2010}
\begin{align}\label{def:work}
\delta w_j :=  a_{i j} X_{j-1} \circ \mathrm{d}X_j ,
\end{align} 
(``coupling force times displacement'')
applied to the subsystem $X_{j}$ via a single interaction between $X_j$ and $X_{i}$.
(Note the usage of Stratonovich calculus throughout the paper.) 
If the coupling is reciprocal, i.e., stems from an interaction Hamiltonian $H_\mathrm{int}$, then the sum of work applied to $X_{i}$ and $X_{j}$
corresponds to its total differentiable
\begin{align}\label{eq:WorkSumConservative}
\delta w_{i} +\delta w_j =-\frac{\partial H_\mathrm{int}}{ \partial X_{i}} \circ \mathrm{d}{X}_{i} -\frac{ \partial H_\mathrm{int}}{ \partial X_j} \circ \mathrm{d}{X}_j =  -\mathrm{d}H_\mathrm{int}  . 
\end{align}
For stability reasons, the latter must on average be conserved in steady states, hence, $\sum_{j=0}^n \dot{ W}_j \equiv 0$.
(In this case,~(\ref{eq:WorkSumConservative}) is actually part of the internal energy of the system.)
On the contrary, a NR coupling is associated with a steady \textit{energy input} into the system at rate
\begin{eqnarray}\label{eq:Einput}
 \dot{E}_\mathrm{input}:=\dot{W}_0+\sum_{j=1}^n \dot{W}_j \geq 0, 
\end{eqnarray}
where equality is reached for symmetric coupling, and the positivity holds for isothermal conditions. (We recall that capital letters denote ensemble averages). 

To prove the non-negativity, we consider the energy balance. 
One shall expect that all picked up energy must ultimately dissipate. The validity of the first law of thermodynamics for each d.o.f: $\delta q_j = \delta w_j + \mathrm{d} u_j$, is indeed already implied in the model~(\ref{eq:Network}), as given by Sekimoto's well-known formalism~\cite{Sekimoto2010}. Besides the fluctuating work $\delta w_j$ defined in~(\ref{def:work}), the first law involves the heat flow 
\begin{align}
\delta q_j = (\gamma_j \dot{X}_j - \xi_j)\circ \mathrm{d}{X}_j = F_j \circ \mathrm{d}{X}_j,
\end{align}
and the internal energies $ \mathrm{d} u_j = a_{jj} X_j \circ \mathrm{d} X_j $. As the average $\langle \mathrm{d} u_j \rangle =\langle a_{jj} X_j \circ \mathrm{d} X_j \rangle =a_{jj} \mathrm{d}\langle X_j^2\rangle =0$ is conserved in steady states, the first law implies $\dot{Q}_j = \dot{ W}_j $. 
Thus, the energy input $\dot{E}_\mathrm{input}$, which we define in~(\ref{eq:Einput}), is indeed identical to the total dissipation rate $\sum_{j=0}^n \dot{Q}_j \geq 0$. The latter is non-negative, as directly follows from the connection to the EP [see (\ref{eq:meanStot_general})]. 
A (positive) energy input on the level of the individual constituents is considered a defining property of active system~\cite{ramaswamy2010mechanics,ramaswamy2017active,nardini2017entropy,fodor2016far,dauchot2019chemical}. Hence, NR networks model active matter. \blue{In accordance with these considerations model (\ref{eq:ExampleI}) was indeed used in the literature to model ...}
 
We note that the energy flow through the system could be initiated by a (bio)chemical energy source, or some sort of external gradient, or force field. In any case, the source will be exploited, such that, looking at the entire system plus surrounding, this steady energy flow must result in free energy loss (or another respective thermodynamic potential). This can be seen by the fact that the energy flow is accompanied by total EP. For this reason, the energy input $\dot{E}_\mathrm{input}$~(\ref{eq:Einput}) can be associated with ``free energy consumption''. 

%

\begin{figure}
\includegraphics[width=0.5\textwidth]{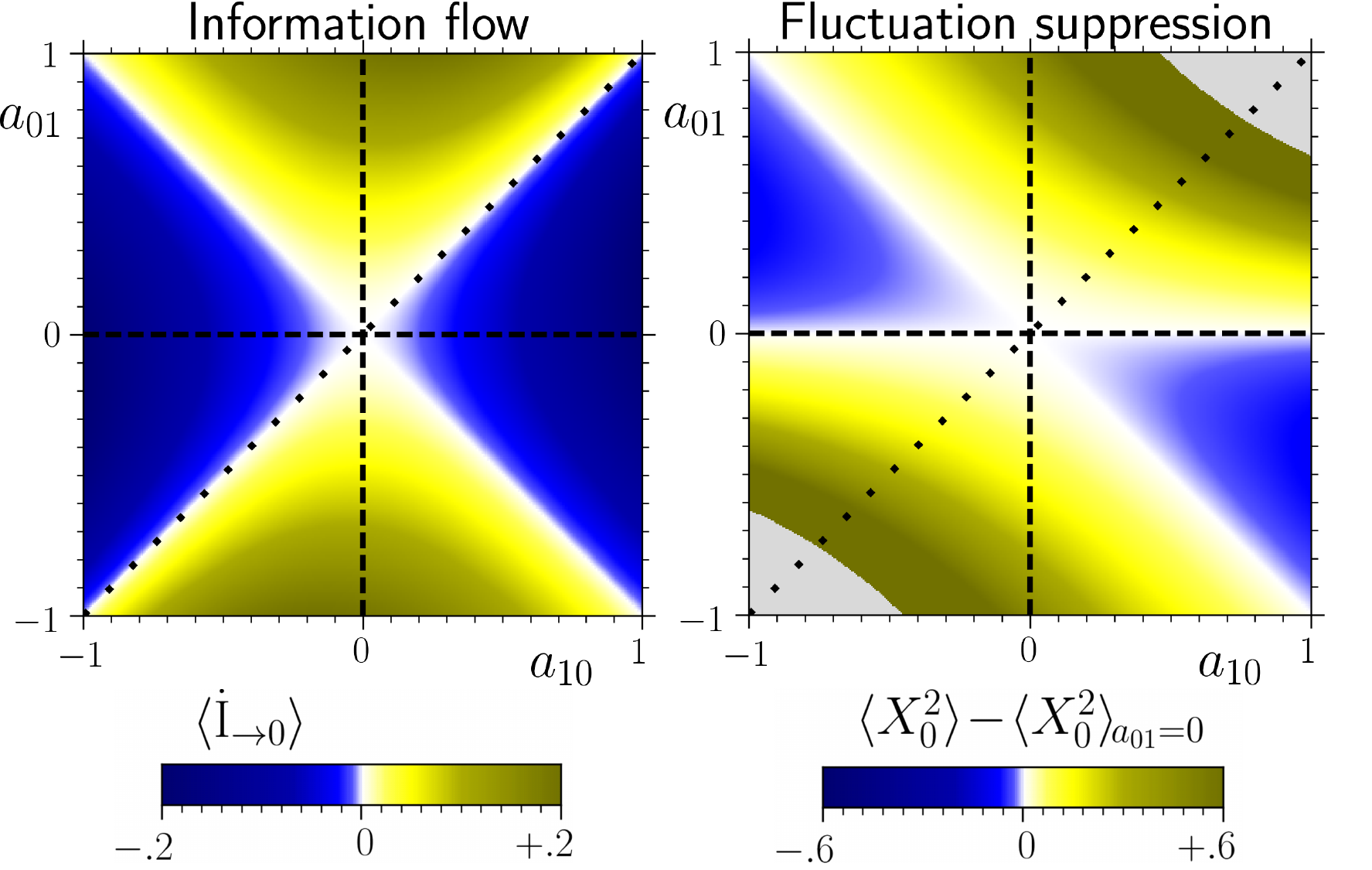} 
\caption{Two d.o.f. $X_0$ and $X_1$, with NR coupling as in Fig.~\ref{fig:heat}. \textit{Left}: Information flow to $X_0$. Blue areas indicate that $X_1$ ``knows'' more about $X_0$ than vice versa corresponding to 
feedback control regimes ($|a_{10}|>|a_{01}|$). \textit{Right:} Thermal fluctuations of $X_0$ measured by the second moment compared to the uncoupled case ($a_{01}=0$), $\langle X_0^2\rangle -\langle X_0^2\rangle_{a_{01}=0}$. Blue areas indicate {thermal fluctuation suppression}. The gray areas 
indicate unstable regions (where $\langle X_0^2\rangle \to \infty$). All other parameters are set to unity and $a_{11} = a_{00} =-1$.
}\label{fig:info}

\end{figure}

Further, the NR networks discussed here also include, as a specific case, a temperature gradient as underlying driving mechanism. 
In particular, if $a_{01}a_{01}\!>\!0$ (upper right and lower left quadrants in Figs.~\ref{fig:heat},\ref{fig:info}), the NR coupling is (potentially) ascribable to a \textit{hidden} temperature gradient, as can be shown by a mapping the network onto a symmetric system %
(see the Appendix~\ref{sec:hiddenTemp} for more details). Hence, this represents a ``mild'' form of intrinsic non-equilibrium. Indeed, (hidden) temperature gradients have been discussed in the literature as possible mechanisms that fuel active motion, see, e.g.,~\cite{roldan2018arrow,netz2018fluctuation}. 
However, for the general case (unidirectional coupling, or $a_{01}a_{01} < 0$), such a mapping cannot be found. Hence, the NR coupling approach taking here is somewhat more general.

\section{NR coupling $\&$ Information} 
%
%
%
%
%
%
%
In the previous section, we have established that NR coupling implies an energy input on the level of individual constituents, which is a defining property of active matter. Now we change the focus onto information and feedback. 

We will address two questions. We have argued that a unidirectional coupling to another d.o.f. can model an active force onto $X_i$ (of strength $a_{ij}$), i.e, ``active propulsion''; or a sensing operation (of strength $a_{ji}$) [see example (\ref{eq:ExampleI})]. 
Do these interpretations conform with the information-thermodynamic properties induced by the unidirectional coupling? 
 
Furthermore, a combination of both, i.e., sensing and applying an (active) force according to the measurement outcome, 
is in line with the intuitive notion of feedback. But is the existence of a single NR coupling indeed sufficient to model a feedback loop? 

We clarify these questions by analyzing information flows and their connection to energy flows~\cite{allahverdyan2009thermodynamic} and entropy~\cite{Seifert2012} in detail. To this end, we use a framework similar to~\cite{allahverdyan2009thermodynamic}, where the reciprocally coupled version case of (\ref{eq:ExampleI}) was studied in detail.

\subsection{Reversed heat flow}
First we revisit the heat flow~(\ref{eq:Q0_n=1}). We recall that in a passive system, e.g., $X_0$ in the uncoupled ($a_{0n}=0$) or the reciprocally coupled case, a non-conservative force acting on $X_0$ can induce a steady-state heat flow. The latter is strictly \textit{nonnegative}, as dictated by the second law of thermodynamics 
$\dot{Q}_0/ \mathcal{T}_0 = \dot{S}_\mathrm{tot}\geq 0$. %
 
Thus, heat is never flowing (on average) from the bath to the system.
Remarkably, the coupled subsystem $X_n$, which also applies a non-conservative force onto $X_0$ (namely $a_{0n} X_n$) \textit{can} induce a {reversed heat flow} $\dot{Q}_0 < 0$ (which is possible due to the usage of extracted information, as we discuss below). This implies a steady extraction of energy from the bath, which is converted into work $\dot{W}_0$, i.e., a (potentially useful) form of energy. %
It is, of course, well-known that such a reversed heat flow can be induced by ``Maxwell-demon'' type of devices \cite{Koski2014,maxwell2001theory}. The system considered here represents a minimal, time-continuous version of such a device.
  
For the case $n=1$, we find that reversed heat flow regimes occur, if $a_{01}a_{10}>0$ (blue regions in Fig.~\ref{fig:heat}). The corresponding feedback force is $a_{01}X_1 \sim a_{01}a_{10}/\gamma_1\int e^{a_{11}(t-t')/\gamma_1} X_0(t')\mathrm{d}t'$. Thus, $a_{01}a_{10}>0$ corresponds to a force directed \textit{away from} the past trajectory, i.e., negative feedback. For larger $n$, we find as well that a reversed heat flow only occurs for negative feedback, see Figs.~\ref{fig:heat_n=1-2}, \ref{fig:vs-n}. 
This even includes the case of discrete delay (which we recover for $n\to \infty$, see Sec.~\ref{sec:non-Monotonic}), as we already reported in an earlier study~\cite{Loos2019}. 

Taking a closer look at Fig.~\ref{fig:heat} reveals that a {reversed heat current} occurs, if the ``sensing'' is stronger than the ``active force'' applied to $X_0$, i.e., $|a_{10}|>|a_{01}|$. The reason for that will become clear when we consider the information flows. 
%

\subsection{Information flow and generalized second law}\label{sec:Info}
%
We start by considering the total temporal derivative of the Shannon entropy~(\ref{eq:Stot_general}), i.e., 
\begin{align}\label{eq:SsysTotal}
 \frac{\dot{s}_\mathrm{sh}}{k_\mathrm{B}}=& \frac{-\partial_t \rho_{n+1} }{\rho_{n+1}}+  \sum_{j=0}^{n} \frac{-\left(\partial_{x_j} \rho_{n+1}\right)\dot{X}_j }{ \rho_{n+1}} . 
%
\end{align} 
In the steady state, the first term vanishes. To calculate the ensemble average of the sum, we utilize the closed, {multivariate} Fokker-Planck equation (FPE) for the $(n+1)$-point probability distribution $\rho_{n\!+\!1}( \underline{x},t )$ of $\underline{x}=(x_0,...,x_n)^T$, whose existence is a key benefit of our approach. [There is in general no (closed) FPE for non-Markovian systems of type~(\ref{eq:LE-X0}).] The FPE reads~\cite{Loos2019b}
\begin{align}\label{eq:FPE-extended}
\partial_t \rho_{n+1} =   -\sum_{j=0}^{n} \partial_{x_j} \underbrace{
 \left[\frac{F_j}{\gamma_j} - \frac{ k_\mathrm{B}\mathcal{T}_j}{\gamma_j }\, \partial_{x_j}\right]\rho_{n +1}}_{=J_j( \underline{x},t )},
\end{align} 
with the probability currents $J_j$. We consider natural boundary conditions $\lim_{x\to \pm \infty}\rho(\underline{x}) =0$, and denote improper integrals $\lim_{r\to\infty} \int_{-r}^{r}$ simply as $\int$. Further, we use $\langle \dot{X}_j A(x_j,t) \rangle = \int J_j A(x_j,t) \mathrm{d}x_j$~\cite{Reimann2002,Seifert2012}.
With these tools, we find the ensemble average of each summand of~(\ref{eq:SsysTotal})
\begin{align}
&\left \langle  \frac{-\left(\partial_{x_j} \rho_{n+1}\right)\dot{X}_j }{\rho_{n+1}} \right \rangle =   
 \int  \frac{-\left(\partial_{x_j} \rho_{n+1}\right) J_j }{  \rho_{n+1}} \, \mathrm{d}\underline{x} 
\nn
=& -\int \underbrace{\left[ \ln(\rho_{n+1}) J_j\right]_{-\infty}^{\infty}}_{\to 0} \,\mathrm{d}\underline{x}_{i\neq j} 
+ \int  \ln \rho_{n+1}(\underline{x})  \partial_{x_j} J_j   \,\mathrm{d}\underline{x} 
\nn
=&-\underbrace{ \int \! \ln \frac{\rho_{1}(x_j)}{  \rho_{n+1}(\underline{x})}\partial_{x_j} J_j \,\mathrm{d}\underline{x} }_{ =  \dot{I}_{\to j} } + \int \ln\rho_1(x_j) \,\partial_{x_j} J_j  \,\mathrm{d}\underline{x} 
\nn
=&- {\dot{I}}_{\to j} + \int \ln\rho_1(x_j)\, \left\{-\partial_{t}\rho_{n+1} +\sum_{i\neq j}\partial_{x_i} J_i  \right\} \,\mathrm{d}\underline{x} 
\nn
=&- {\dot{I}}_{\to j} + \frac{\dot{S}^j_\mathrm{sh}}{k_\mathrm{B}}-  \sum_{i\neq j}\int \ln \rho_1(x_j) \, \underbrace{\left[  J_i \right]_{-\infty}^{\infty}}_{\to 0} \,\mathrm{d}\underline{x}_{l\neq i},
\label{eq:Iflow}
\end{align}
where we have introduced the change of the Shannon entropy of the \textit{marginal} pdf $\rho_{1}(x_j)$: $\dot{S}^j_\mathrm{sh}=-k_\mathrm{B}\int \!  \ln   \rho_{1}(x_j) \partial_{t} \rho_{1}(x_j)\, \mathrm{d}{x_j}$,
as well as the information flow ${\dot{I}}_{\to j}$ to $X_j$.
We stress that the involved information flow is from \textit{all} other d.o.f. $\{X_{l\neq j}\}$ to $X_j$. (Even if not directly coupled with each other, two d.o.f. can exchange information through a third d.o.f.) 
It is closely connected to the mutual information between all d.o.f.,
\begin{align}\label{def:mutualInfo}
\mathcal{{I}}(\underline{x}) =\int \rho_{n+1}(\underline{x}) \ln \frac{  \rho_{n+1}(\underline{x})}{ \rho_{1}(x_0)\rho_{1}(x_1)..\rho_{1}(x_n) }  \,\mathrm{d}\underline{x},
\end{align}
via $\dot{\mathcal I}=\sum_{j=0}^{n} I_{\to j}$, see the Appendix~\ref{sec:AppInfo} for details. Since the latter itself is a conserved quantity in the steady state, i.e., $\mathcal{\dot{I}} =0$, the information flows among all subsystems in total cancel each other out (thus, from an information-theoretical point of view, the network as a whole is ``closed''). 
However, they are an important contribution to the entropy balance, when an \textit{individual }subsystem is considered.

To see this, we again consider the summands of~(\ref{eq:SsysTotal}), and rewrite them using the FPE~(\ref{eq:FPE-extended}) as
\begin{align}
\frac{-\left(\partial_{x_j} \rho_{n+1}\right)\dot{X}_j }{k_\mathrm{B}^{-1} \rho_{n+1}}= \underbrace{ \frac{\gamma_j \,J_j( \underline{x},t )\dot{X}_j }{\mathcal{T}_{j}\,\rho_{n+1}} }_{:=\dot{s}_\mathrm{tot}^j/k_\mathrm{B}}- \frac{\dot{q}_j }{\mathcal{T}_{j}}  . \label{eq:STotalJ}
\end{align}  
Combining (\ref{eq:SsysTotal}, \ref{eq:Iflow}, \ref{eq:STotalJ}) shows that a definition of a fluctuating ``total EP per subsystem'', $\dot{s}_\mathrm{tot}^j $,
is meaningful, as it has the non-negative ensemble average
\begin{align}
\dot{S}_\mathrm{tot}^j & = \dot{S}^j_\mathrm{sh}  -k_\mathrm{B}{\dot{I}}_{\to j} + \frac{\dot{Q}_j}{\mathcal{T}_j} =
 \int \frac{\gamma J_j( \underline{x},t )^2}{\mathcal{T}_{j}\,\rho_{n+1}}  \,\mathrm{d}\underline{x}\geq 0 , \label{eq:Sj1}
\\
 \dot{S}_\mathrm{tot} & = \sum_{j=0}^{n} \frac{\dot{Q}_j}{\mathcal{T}_j} + \dot{S}_\mathrm{sh}=\sum_{j=0}^{n} \dot{S}_\mathrm{tot}^j   \geq 0 . \label{eq:Sj3}
\end{align}
With~(\ref{eq:Sj3}) we have recovered the mean total EP~(\ref{eq:Stot_general}) deduced from the path probabilities.

Further, Eq.~(\ref{eq:Sj1}) is a generalized second law for each d.o.f.. As for steady states $\dot{S}^j_\mathrm{sh}=0$, it simplifies to
\begin{align}\label{eq:GeneralizedSecondLaw}
{\dot{Q}_j} \geq k_\mathrm{B}\mathcal{T}_j\,{\dot{I}}_{\to j},
\end{align}
in accordance with~\cite{horowitz2014second,allahverdyan2009thermodynamic}. 

\begin{figure}
\includegraphics[width=0.38\textwidth]{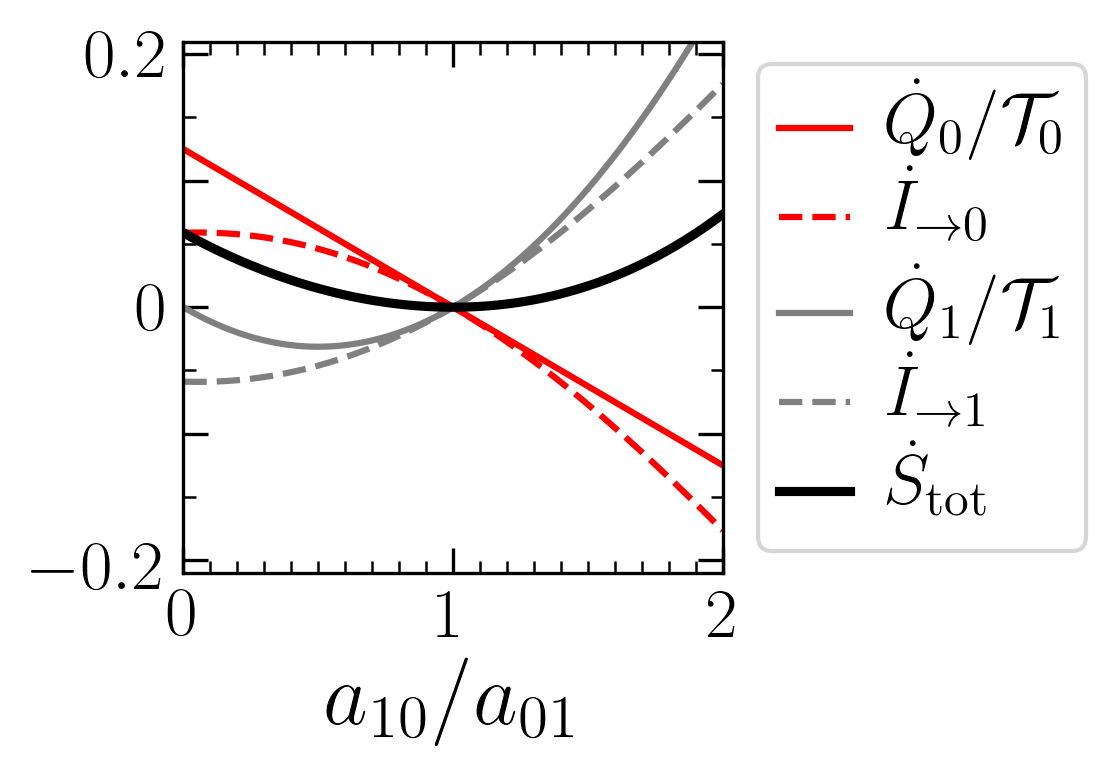} 
\caption{Information $\dot{I}_{\to j}$~(\ref{eq:Info-flow-0}) and heat $\dot{Q}_{j}$~(\ref{eq:Q0_n=1}) flows for two NR coupled d.o.f., vs. the ratio of coupling strength, $a_{01}/a_{10}$. At $a_{01}=a_{10}$, the system is reciprocally coupled and displays no net heat and information flow, and has zero EP (black line). The heat flow (solid red and gray lines) of each d.o.f. is bounded from below by the information flow (dashed lines), as predicted by the generalized second law~(\ref{eq:GeneralizedSecondLaw}). The total EP is given by the sum over all other quantities. $a_{11} = a_{00} =-1$, all other parameters and $k_\mathrm{B}$ are set to unity.
}\label{fig:Info+Heat}
\end{figure}
%
%
%

Equation~(\ref{eq:GeneralizedSecondLaw}) states that a reversed steady heat flow, $\dot{Q}_0<0$, is only possible, if $\dot{I}_{\to 0}<0$, i.e., information is flowing from the $X_0$ to the rest of the system. The more information about $X_0$ is gathered by the other $X_{j>0 }$ (the controller d.o.f., see Sec.~\ref{sec:non-Monotonic}), the more heat can be extracted from the bath. Figure~\ref{fig:Info+Heat} shows [for model~(\ref{eq:ExampleI})] the information and heat flows, as well as the total EP, which are all connected via~(\ref{eq:Sj1}, \ref{eq:Sj3}). It also illustrates that in the symmetric case, the memory is free of ``entropic cost'' (zero EP), but, at the same time, no net information extraction is achieved, nor is a heat flow induced.


Due to the linearity of the model, the steady-state pdfs are multivariate Gaussians with zero mean and with the covariance matrix $(\underline{\underline{\Sigma}})_{i j}=\langle X_i X_j\rangle$, as already mentioned in Sec.~\ref{sec:analytical_solutions}. To derive explicit expressions for the steady-state averaged information flows, it turns out to be most convenient to use from~(\ref{eq:Iflow}) [recall $\dot{S}^j_\mathrm{sh}=0$]
\begin{align}\label{eq:generalFormula_Infoflow0}
{\dot{I}}_{\to j} 
=&\left \langle \frac{\left(\partial_{x_j} \rho_{n+1}\right)\dot{X}_j }{\rho_{n+1}} \right \rangle
= -\left \langle(\underline{\underline{\Sigma}}^{-1}\underline{X})_j \dot{X}_j \right \rangle.
\end{align}
In the last step, we have utilized a general property of normal distributions $\partial_{x_j} \rho_{n+1}(\underline{x})= $ $ - (\underline{\underline{\Sigma}}^{-1}\underline{X} )_j\rho_{n+1}$. Substituting~(\ref{eq:Network}), we find the general formula 
\begin{align}\label{eq:generalFormula_Infoflow}
{\dot{I}}_{\to j} 
&= - \sum_{l\neq j}   (\underline{\underline{\Sigma}}^{-1})_{lj} \left[  \frac{a_{jj} }{\gamma_j} \langle {X}_l {X}_j \rangle +\frac{a_{j j-1}}{\gamma_j}  \langle {X}_l X_{j-1}\rangle \right].
\end{align}
In combination with~(\ref{eq:Solution-Corr}), Eq.~(\ref{eq:generalFormula_Infoflow}) represents a closed expression for the information flow to any node in the network of arbitrary size.  
The explicit expressions are, however, quite cumbersome. As an example, the result for $n=2$ is given in the Appendix, Eq.~(\ref{eq:info_n=2}).
For $n=1$, the information flow to $X_0$ simplifies to
\begin{align}\label{eq:Info-flow-0}
\mathcal{T}_0 {\dot{I}}_{\to 0} =  \frac{ -\langle X_1 X_0\rangle }{\langle X_1 X_0\rangle^2  -\langle X_1^2\rangle \langle X_0^2\rangle }\frac{\dot{ Q}_0}{\gamma_0\,a_{01}}\nn
=
\frac{ [a_{01} \mathcal{T}_1-a_{10} \mathcal{T}_0][ a_{00} a_{01}/\mathcal{T}_0 +
 a_{11} a_{10} \gamma_0/(\mathcal{T}_1\gamma_1)]
 }
{ (a_{00} \gamma_1+a_{11}
\gamma_0)^2+  \frac{a_{01}^2\mathcal{T}_1}{\mathcal{T}_0\gamma_0 \gamma_1}  -2 a_{01} a_{10} +a_{10}^2  \frac{\mathcal{T}_0}{\mathcal{T}_1} }.
\end{align}
According to~(\ref{eq:Info-flow-0}), the information flow trivially vanishes if $X_0$ and $X_1$ are uncorrelated as one would expect. It also vanishes for reciprocally coupled systems (or, if $\mathcal{T}_0 a_{01}=\mathcal{T}_1 a_{10}$ at non-isothermal conditions), i.e.., when the subsystems reach thermal equilibrium. Furthermore, ${\dot{I}}_{0\to }=0$
if $\dot{Q}_0=0$, in accordance with~\cite{allahverdyan2009thermodynamic} (but only if $a_{01}\neq 0$).

We are now in the position to clarify the meaning of unidirectional coupling from an information-thermodynamic perspective. Inspecting Figs.~\ref{fig:heat} and \ref{fig:info} we see that along the unidirectional coupling axis $a_{01}=0$, there is net information flow from $X_0$ to $X_1$, but no net work applied onto $X_0$ ($\dot{Q}_0=\dot{W}_0=0$). 
Thus, it is indeed very reasonable to consider $X_1$ a ``sensor'' and the unidirectional coupling a \textit{sensing interaction}. On the other hand, if the unidirectional coupling is reversed ($a_{10}=0$), the heat flow is always positive, $\dot{Q}_0>0$. This suits the idea that $X_0$ models an active swimmer within the AOUP model. In particular, the swimmer heats up the surrounding fluid but never has a net cooling effect, as shall be expected. In this case, there is as well a nonzero information flow, which is directed from the source of propulsion (i.e., the flagella or flow field) to the particle. This is indeed reasonable as the propulsion force ``carries'' information (one could, on average, reconstruct the position of the flagella by only monitoring $X_0$). 
%

\subsection{Feedback}
According to the results from the last section, a (unidirectional) interaction from one d.o.f., say $X_0$, to another, $X_1$, can model a measurement operation, while the backwards coupling can model an active (propulsion) force. Thus, a NR interaction can be considered as a combination of a measurement {and} a force depending on the measurement outcome, or, in other words, \textit{feedback}. In this sense, we can state that our networks with NR non-unidirectional coupling model feedback loops. 

However, it seems intuitive to consider $X_0$ a feedback-\textit{controlled} system, if and only if the net information flow out of $X_0$ is {positive}, i.e., the controller ``knows'' more about $X_0$ then vice versa.

According to this definition, the \textit{control regime} for the model (\ref{eq:ExampleI}), is given if $|a_{10}|>|a_{01}|$, see the blue regions in Fig.~\ref{fig:info}. The generalized second law~(\ref{eq:GeneralizedSecondLaw}) states that only in this regime, a reversed heat flow is possible, as indeed confirmed by Fig.~\ref{fig:heat}. Interestingly, we find that another intricate phenomenon is limited to the control regime, namely, the \textit{suppression of thermal fluctuations} of $X_0$, measured by a reduced second moment $\langle X_0^2 \rangle < \langle X_0^2 \rangle_{a_{01}=0}$ (see Fig.~\ref{fig:info}). Such a suppression, i.e., noise-reduction (which can be interpreted as feedback cooling) is desired in various experimental setups, and indeed one of the main applications of feedback control~\cite{Steck2006,Cohadon1999,Vinante2008}. 

We thus conclude that characteristic features of a feedback controlled systems are described by a single NR coupling, e.g., by~(\ref{eq:Network}) with $n=1$ and $|a_{10}|>|a_{01}|$. $X_1$ represents a controller that utilizes information about $X_0$ extracted via a sensing operation. The resulting system is characterized by positive information flow from $X_0$ to $X_1$, reversed heat flow, and suppression of thermal fluctuations (depending on the strength of the ``feedback force'' $a_{01}X_1$).

For further discussions about information flow, energetics and control, we refer the interested reader to~\cite{allahverdyan2009thermodynamic}.

\section{NR coupling $\&$ Non-monotonic memory}\label{sec:non-Monotonic}
So far, we have established that NR coupling alone is sufficient to model active, feedback-controlled systems. 
While the presented formulae are more general, we have, so far, mainly focused on $n=1$, i.e., an exponentially decaying memory kernel [model (\ref{eq:ExampleI}) and model (\ref{K_noise_II}) with $n=1$]. However, many real-world systems involve non-monotonic memory. Here we discuss memory kernels with a peak around a non-zero delay time $\tau$, which
are a characteristic of delayed feedback control.

\subsection{Emergence of non-monotonic memory}
As a first step, we 
show that 
already very small networks with NR coupling can model non-monotonic memory.
%
\begin{figure}
\includegraphics[width=0.17\textwidth]{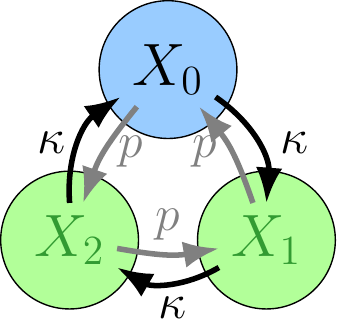} 
\caption{Ring network of three nodes. 
For symmetric coupling $\kappa=p$, the network corresponds to a model for three colloids  reciprocally coupled by springs of stiffness $\kappa$. In this case, the memory kernel is exponentially decaying, while NR coupling yields non-monotonic memory, see Eq.~(\ref{eg:kernel_n=3}).}\label{fig:nonmonotonic}
\end{figure}
Let us consider as an illustrative example a ring network~(\ref{eq:Network}) of three nodes (i.e., $n=2$,
 $a_{j j-1} =\kappa$, $a_{j j+1}= p$, $a_{jj} =-(p+\kappa)$), 
as schematically shown in Fig~\ref{fig:nonmonotonic}.
This ring network generates the memory kernel
\begin{align}\label{eg:kernel_n=3}
K(t-t')=
\frac{e^{\left(-\sqrt{p \kappa}+p+\kappa\right) t'-t \left(\sqrt{p \kappa}+p+q\right)}}{2 \sqrt{p \kappa}}\times
\nn
\left[ \left(p^{3/2}+\kappa^{3/2}\right)^2 e^{2 \sqrt{p \kappa} t}
-\left(p^{3/2}-\kappa^{3/2}\right)^2 e^{2 \sqrt{p \kappa} t'}\right],
\end{align}
(see the Appendix~\ref{sec:MemoryRing} for a derivation).

For \textit{reciprocal} coupling $\kappa=p$, this network can be considered as a model for three overdamped particles connected by harmonic springs with spring constant $\kappa$, which has the corresponding Hamiltonian $H=(\kappa/2) \sum_{i\neq j} (X_i-X_j)^2$. In this case, (\ref{eg:kernel_n=3}) simplifies to an exponential decay, $K(T) =2 \kappa^2 e^{-\kappa T} $. We observe the same qualitative behavior for larger networks with $n>2$. 

In any other case, $\kappa\neq p$, the coupling is stronger in one direction of the ring than in the other direction. Then, the memory becomes non-monotonic. For example, in the limit of unidirectional coupling $p\to 0$, where this model becomes a specific realization of model (\ref{K_noise_II}), the memory kernel~(\ref{eg:kernel_n=3}) converges to $K(T)= \kappa^3 T e^{- \kappa T }$, as displayed in Fig.~\ref{fig:kernels} (b). 

Playing around with different coupling topologies, we generally find that reciprocal couplings yield monotonic memory kernels. We thus conclude that NR coupling is a crucial ingredient to generate non-monotonic memory. In the following, we will consider model (\ref{K_noise_II}), which is essentially a generalization of this unidirectional ring to arbitrary lengths. Before we proceed with exploring the thermodynamic properties, we discuss a possible interpretation of the d.o.f. $X_{j>0}$.
\subsection{Interpretation of $X_{j>0}$ in the case of feedback control}
%
%
%
%
%
\begin{figure}
\includegraphics[width=0.3\textwidth]{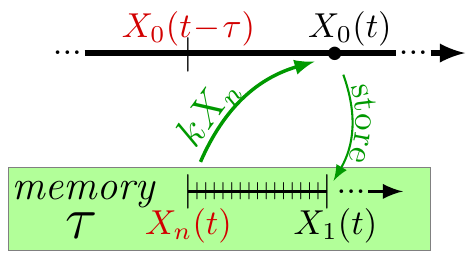} 
\caption{Relation between the temporal evolutions of the ``system'' $X_0$ and the $n$ memory cells $X_{j>0}$ of the controller.}\label{fig:Sketch_Memory}
\end{figure}
%
%
%
%
Let us take a closer look at an important example of higher dimensional networks, that is, feedback control of colloidal particles via optical tweezers. Here, the memory kernel is typically sharply peaked around a delay time $\tau$.%

In experimental setups, the delay is typically created within a computer's memory, storing the past trajectory of length $\tau$. In this case, the additional d.o.f. in the network~(\ref{K_noise_II}) can be interpreted as the memory cells of the controller. 

More specifically, imagine a memory device with a ``shift register'' type of architecture, as sketched in Fig.~\ref{fig:Sketch_Memory}. At every instant in time, the information stored in each cell $X_j$ is shifted one step further from $j$ to $j+1$, while the new information (gathered by a measurement operation) is
stored in the first cell $X_1$. Thus $X_n$ stores the oldest information and is used to perform the feedback control force $k X_n$.
Mathematically, this operation can be represented in discretized form as $\Delta  X_j = \frac{\Delta t}{\tau/n} [X_{j-1}-X_j] + \epsilon_j$, where $\epsilon_j$ is a random number accounting for errors, e.g., from measurement, rounding operations, or a finite temperature of the device.
The updates of the memory cells are scaled with $\tau/n$ to enable storage of a trajectory of length $\tau$ in $n$ cells.
Since we aim to model time-\textit{continuous }feedback loops, we assume that the updates are performed (infinitely) fast, that is, $\Delta t\to 0$, yielding the {dynamical} equations of ${X}_j$ in~(\ref{eq:Network}). For sake of simplicity, we set $\mathcal{T}_{j>0}=\mathcal{T}'$. In the present context this implies the magnitude of error is identical in each memory cell. 

As suggested in~\cite{Horowitz2017}, such a memory device could be realized by $n$ colloidal particles at positions $X_j$, each trapped in a harmonic trap of stiffness $n/\tau$ centered around the position of the proceeding colloid, $X_{j-1}$. This idea is, of course, rather an idealized thought experiment. Still, it is useful to illustrate how a delay, i.e., a memory kernel concentrated around a specific instant in time of the past (the delay time), can result from coupled exponential relaxation processes. It further demonstrates that an implementation of (non-monotonic) memory, with physical d.o.f. is inevitably associated with thermodynamic cost and dissipation, which we will discuss below. ({Interestingly, this thought experiment also illustrates that the friction terms $\gamma'\dot{X}_i $ may be crucial to memorize, because they enable ``forgetting'' the old information.})

In line with the examples used in the first part of this paper, such a ``memory device'' could also be a cascade of chemical reactions. In fact, in~\cite{hartich2016sensory} a model for a cellular sensor ($X_1$) with one ``memory cell'' ($X_2$) is proposed, which exactly corresponds to our model with $n=2$ and $k=0$ (i.e., no backcoupling from $X_2$ to $X_0$, since~\cite{hartich2016sensory} doesn't consider feedback). The focus of~\cite{hartich2016sensory} was to calculate information flows for $n=1,2$. 
We recall~(\ref{eq:GeneralizedSecondLaw}), which states that the information flow to $X_0$ provides a lower bound to the heat flow between $X_0$ and its bath.
We will here focus on other thermodynamic quantities, particularly the heat and total EP, and study how they depend on the number of d.o.f., $n$.
%
%
%
%
\subsection{Thermodynamics in the presence of non-monotonic memory}
%
%
%
%
\begin{figure}
\includegraphics[width=0.5\textwidth]{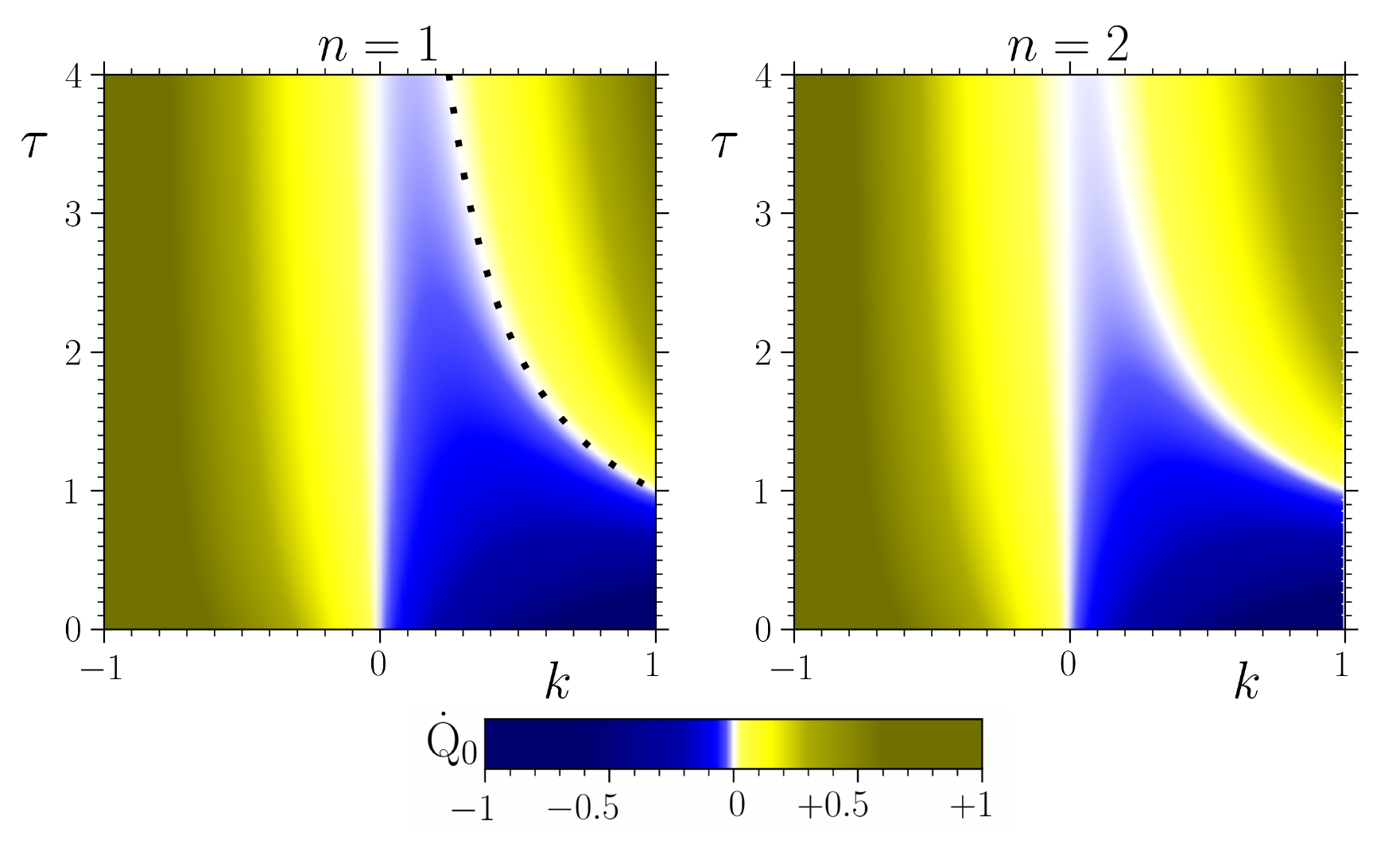} 
\caption{Heat flow from $X_0$ to its bath [from Eqs.~(\ref{eq:Q0_general}, \ref{eq:Q0_n=1})], in networks of type (\ref{K_noise_II}) of different sizes; \textit{left}: $n=1$ (one ``memory cell'', exponential memory) and \textit{right}: $n=2$ (two ``memory cells'', non-monotonic memory). For $n=1$, the FDR~(\ref{eq:Fulfill-FDR}) is fulfilled at $k= 1/\tau$ (dotted line). Since this system equilibrates, $\dot{Q}_0=0$. In the uncoupled case ($k=0$), the subsystem $X_0$ equilibrates as well (for arbitrary $n$). All other parameters are set to unity.}\label{fig:heat_n=1-2}
\end{figure}
%
%
%
%
%
\begin{figure}
\includegraphics[width=0.5\textwidth]{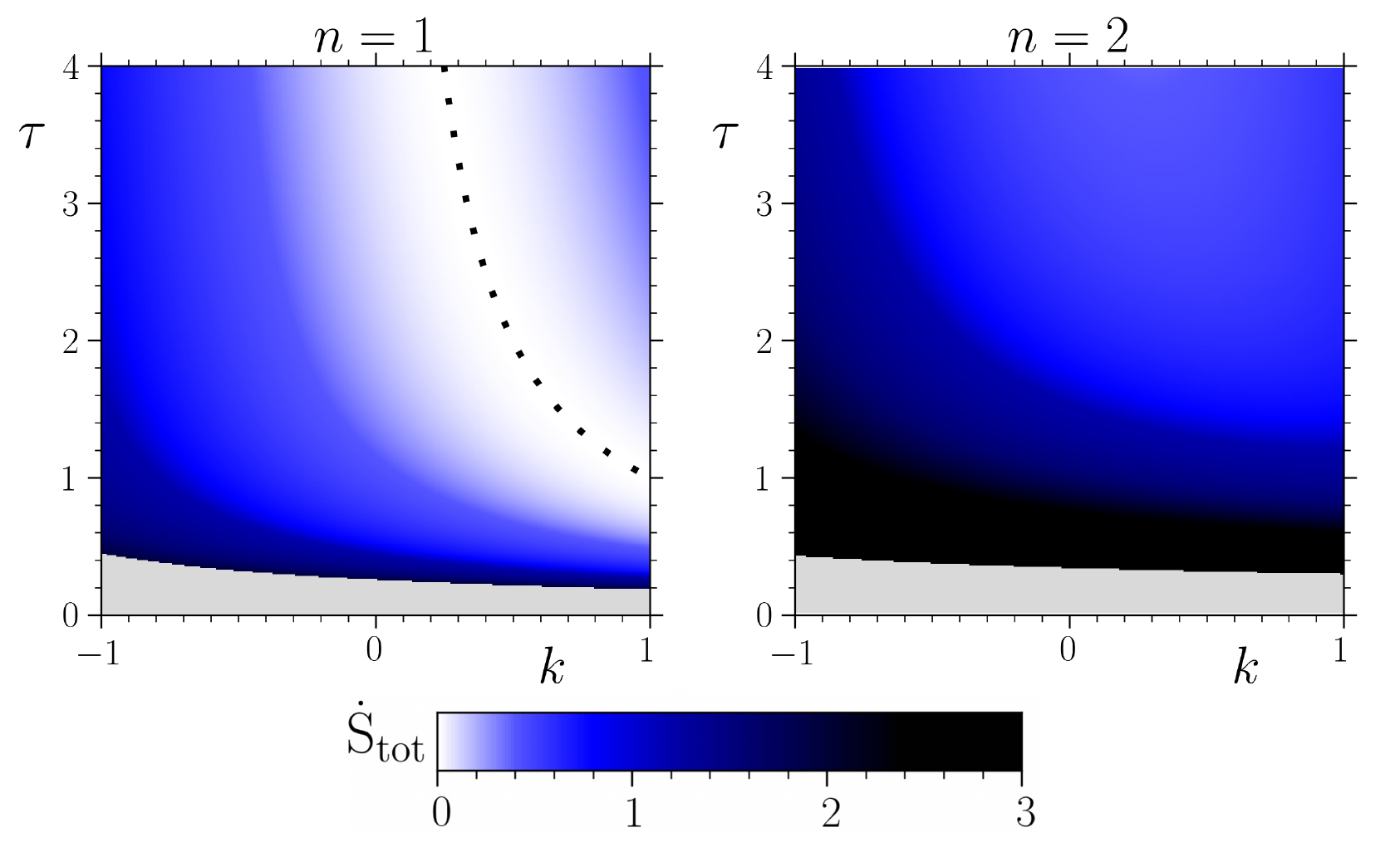} 
\caption{Total EP $\dot{S}_\mathrm{tot}$ [from Eqs.~(\ref{eq:meanStot_general}, \ref{eq:Stot_n=1})], in networks of sizes $n=1,2$, as in Fig.~\ref{fig:heat_n=1-2}. 
Along the dashed line, FDR~(\ref{eq:Fulfill-FDR}) is fulfilled, and the total EP is zero, showing consistency between the non-Markovian description~(\ref{eq:LE-X0}) and the Markovian network description~(\ref{eq:Network}). The gray areas indicate unstable regions (it is slightly larger in the case $n=2$).
}\label{fig:entropy_n=1-2}
%
%
\end{figure}
We revisit the formulae derived in the previous sections.
First, we consider the mean heat flow from $X_0$ to its bath from Eq.~(\ref{eq:Q0_general}), which is given by
$
\dot{Q}_\mathrm{0}  = k^2 \langle  X_n^2\rangle+k a_{00} \langle  X_0 X_{n}\rangle ,
$
with $k$ being the strength of the feedback $k {X_n} $, and $a_{00}X_0$ being the restoring force. 
A comparison of the cases $n=1,2$ is instructive, since the networks are of comparable size while the memory kernels $K(T)$ have distinct characteristics, i.e., exponential decay with maximum at $T=0$ ($n=1$) vs. a non-monotonic kernel with minimum $K=0$ at $T=0$ ($n=2$), see Fig.~\ref{fig:kernels} (a,b). Figure~\ref{fig:heat_n=1-2} reveals that the heat flow $\dot{Q}_0$ is qualitatively and quantitatively similar for $n=1$ and $2$. The (blue) area of reversed heat flow is slightly smaller for $n=2$, but lies in the same region of the $(\tau,k)$-plane. For $k=0$, there is trivially no heat flow (in this case, $X_0$ is a passive d.o.f. which does not ``see'' the other d.o.f.). For both $n=1,2$, there is a second line along which the heat flow vanishes. For $n=1$, this line corresponds to parameters where the FDR~(\ref{eq:Fulfill-FDR}) is fulfilled (dashed line). As shown in Fig.~\ref{fig:entropy_n=1-2}, the total EP is then zero as well. For $n=2$, the FDR is broken for all $(\tau,k)$ and the system never reaches thermal equilibrium. Correspondingly, the EP is strictly positive, 
quantifying the ``thermodynamic cost'' of the non-monotonic memory, or, the dissipation of $X_{j>0}$.
%
\begin{figure}
\includegraphics[width=0.42\textwidth]{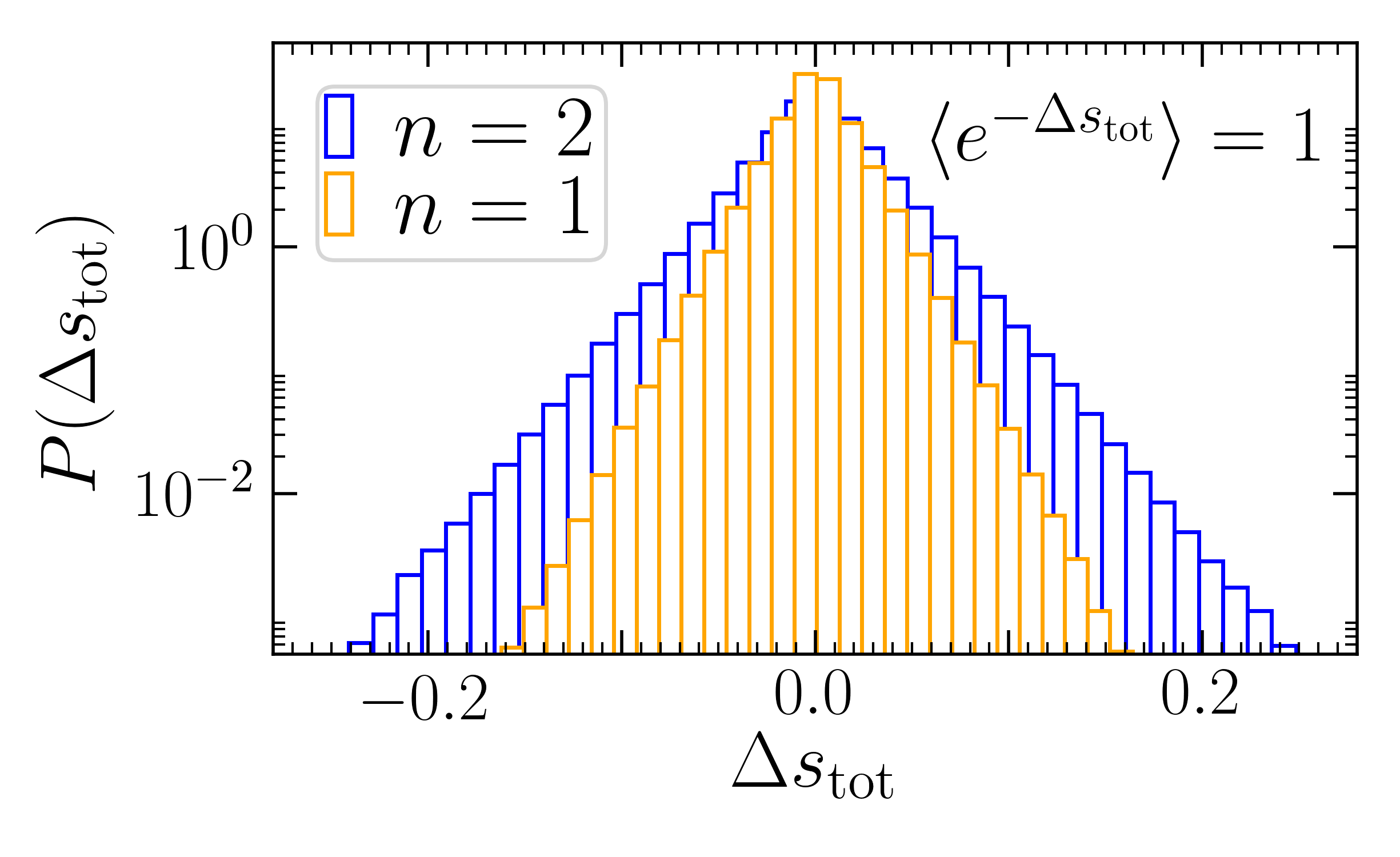} 
\caption{Distributions of the fluctuating total EP $\Delta {s}_\mathrm{tot}$~(\ref{eq:Stot_general}) for $n=1,2$ from Brownian dynamics simulations. The linear decays in this logarithmic plots indicate exponential tails. The mean values are in good agreement with the analytical results ($\Delta S_\mathrm{tot}= 2$ for $n=1$, and $\Delta S_\mathrm{tot}= 5$ for $n=2$), and both distributions fulfill the integral fluctuation theorem~(\ref{eq:IFT}). To calculate the distributions, $> 5\cdot 10^6$ steady-state trajectories of length $10^{-4}$ were generated. $k=-1$, and all other parameters and $k_\mathrm{B}$ are set to unity. %
}\label{fig:Distributions}
\end{figure}

So far, we have focused on the mean values of the total EP, which we have analytically calculated. 
Finally, Fig.~\ref{fig:Distributions} displays (numerically obtained) distributions of the total EP fluctuations.
The distributions have very similar characteristics for $n=1$ and~$2$, despite the distinct memory kernels in both cases (exponential decay vs. non-monotonic memory). In particular, they both have exponentially decaying tails, and fulfill the symmetry $\langle e^{-\Delta s_\mathrm{tot}/k_\mathrm{B}}\rangle =1$, as expected [see discussion below Eq.~(\ref{eq:Stot_general}) and Appendix~\ref{sec:IFT}]. 
The characteristics of the entropy fluctuations are, in fact, identical for all cases, whether the network models an active microswimmer, a cellular sensor, or a feedback-controlled colloid with time delay.
%
%
\begin{figure}
\includegraphics[width=0.5\textwidth]{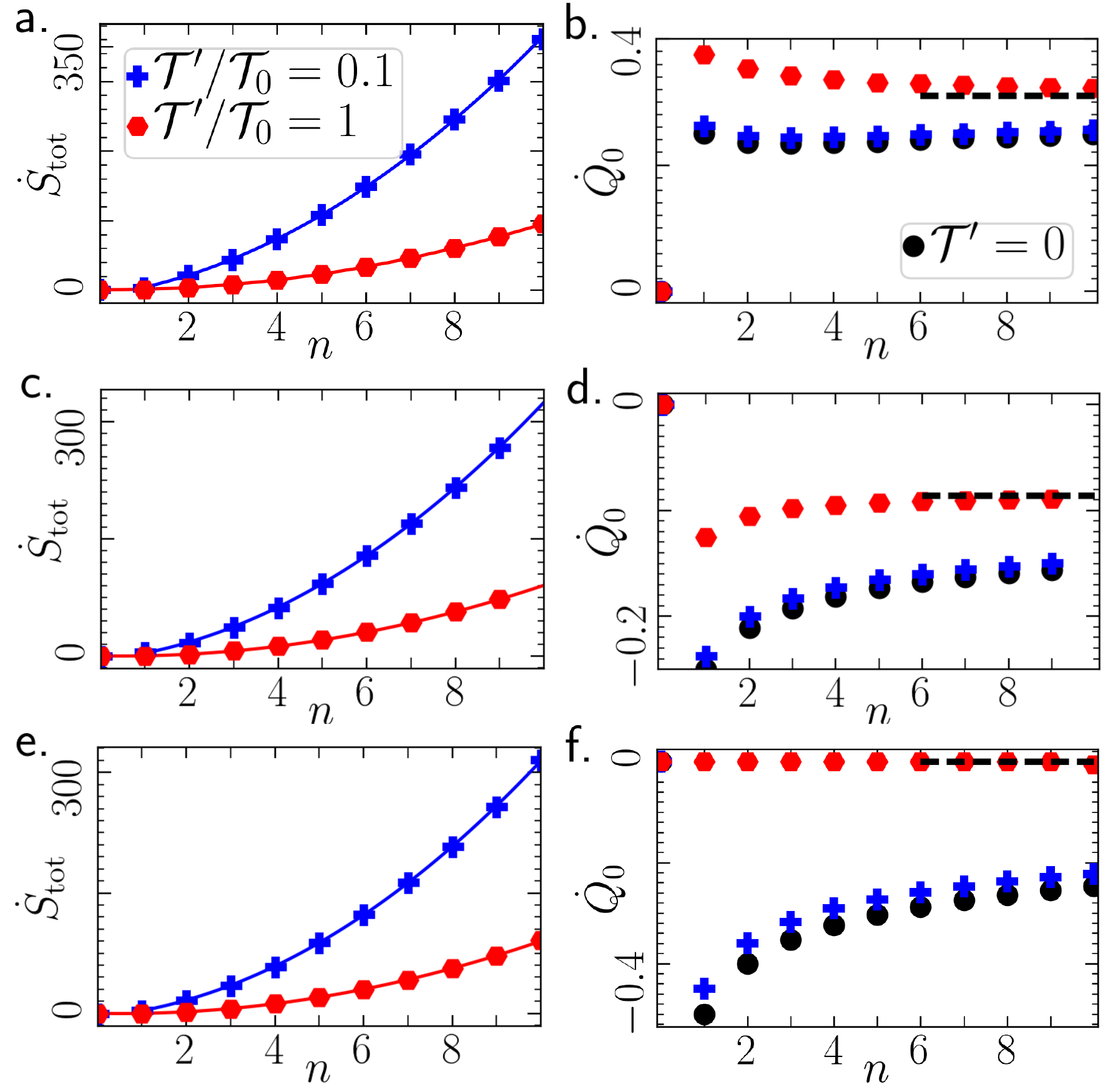}
\caption{(a.+c.+e.) Total EP [from Eq.~(\ref{eq:meanStot_general})], and (b.+d.+f.) heat flow $\dot{Q}_0$ [from Eq.~(\ref{eq:Q0_general})], vs. the number $n$ of $X_{j>0}$. The EP plots are complemented by quadratic fits $\dot{S}_\mathrm{tot}\sim n^2$ (solid lines). For all temperature ratios $\mathcal{T}'/\mathcal{T}_0$ (different colors), the heat flow converges in the limit $n\to\infty$ to the discrete delay solution (dashed lines). At $\mathcal{T}'\!=\!0$ (black disks) $\dot{S}_\mathrm{tot}$ diverges. (a.+b.) $k=1/2$ (positive feedback pointing away from the past trajectory), (c.+d.) 
$k=1/2$ (negative feedback). (e.+f.) $k=1$, where the system with $n=1$ is reciprocally coupled. All other parameters and $k_\mathrm{B}$ are set to unity.
}\label{fig:vs-n}
\end{figure}
%
%
%

Now we further increase the number of memory cells $n$, corresponding to more sharply peaked memory kernels appearing in the dynamics of $X_0$. Figure~\ref{fig:vs-n} shows analytical results for the mean total EP and heat flow as functions of $n$, for three different values of the feedback strength $k$, i.e., the coupling between $X_n$ and $X_0$. 
We observe that the total EP generally increases quadratically with $n$. This observation is robust against the details of the system, e.g., the temperatures or delay time. In fact, it appears to hold even for nonlinear cases~\cite{note3}.

In sharp contrast, the heat flow saturates with $n$, as shown in the right panel of Fig.~\ref{fig:vs-n}. Moreover, as in the case (\ref{eq:ExampleI}) displayed in Fig.~\ref{fig:heat}, a reversed heat flow ($\dot{Q}_0<0$) is found for negative feedback, i.e., when the feedback force is directed away from the past trajectory. 
Finally, Fig.~\ref{fig:vs-n} also indicates that the heat flow converges to the previously obtained result as $n\to \infty$ (dashed lines)~\cite{Loos2019}. Remarkably, it approaches this limit fastest for isothermal conditions $\mathcal{T}_0=\mathcal{T'}$ (see also Fig.\,\ref{fig:vs-T/T}).
We will discuss both observations in more detail below.

%
%
%

\subsection{Impact of measurement errors}

\begin{figure}
\includegraphics[width=0.4\textwidth]{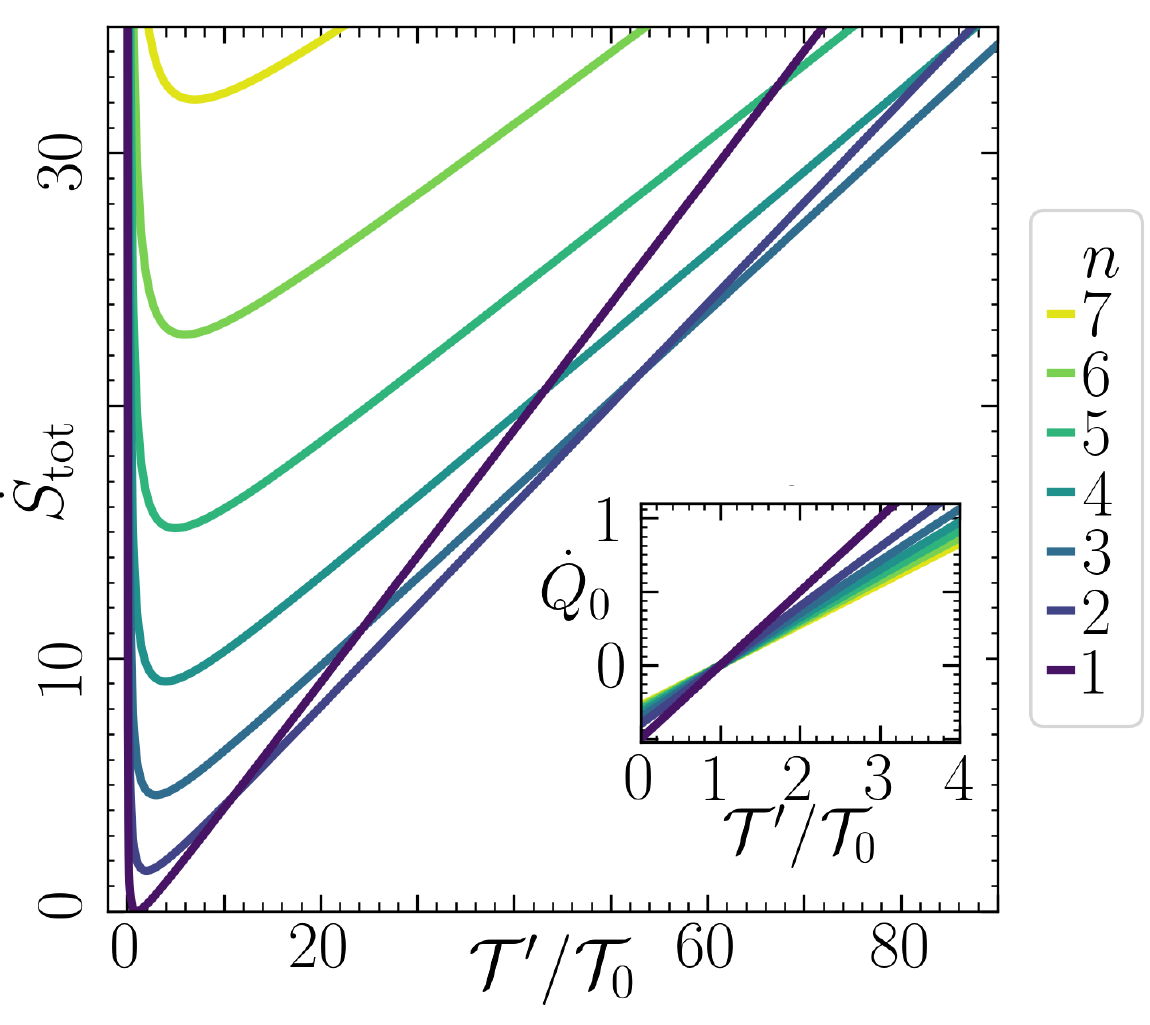} 
\caption{Total EP $\dot{S}_\mathrm{tot}$ [from Eq.~(\ref{eq:meanStot_general})] as a function of the ratio of the temperatures of $X_{j>0}$ and $X_0$, for networks with different numbers $n$ of d.o.f. $X_{j>0}$. The parameters are identical to Fig.~\ref{fig:vs-n} (e.+f.). The network with $n=1$ is reciprocally coupled, and fulfills FDR~(\ref{eq:Fulfill-FDR}) if $\mathcal{T}'/\mathcal{T}_0=1$. In accordance, the total EP vanishes. For each $n$, the EP has a minimum at a finite $\mathcal{T}'/\mathcal{T}_0$ of order $1$, and diverges in the limits $\mathcal{T}'/\mathcal{T}_0\to 0$ or $\mathcal{T}'/\mathcal{T}_0 \to \infty$. 
The inset shows the heat flow $\dot{Q}_0$, which for all $n$ increases linearly and is zero at $\mathcal{T}'/\mathcal{T}_0=1$.
}\label{fig:vs-T/T}
\end{figure}

Here we consider the impact of $\mathcal{T'}$, that is the strength of the noise terms $\xi_j$ in Eq.~(\ref{eq:Network}) (accounting for measurement or rounding errors). 
Figure~\ref{fig:vs-T/T} shows that the heat flow $\dot{Q}_0$ increases linearly with $\mathcal{T'}/\mathcal{T}_0$. Thus, larger measurement errors generally imply an increased heat flow. Due to the specific parameter choice in Fig.~\ref{fig:vs-T/T} (where $n=1$ corresponds to a reciprocally coupled network), the heat flow vanishes at $\mathcal{T}_0=\mathcal{T'}$. 

On the contrary, the total EP has for each $n$ a pronounced minimum at a finite $\mathcal{T'}/\mathcal{T}_0$. Interestingly, at some values of $\mathcal{T'}/\mathcal{T}_0$ above the minima, the total EP for different $n$ are \textit{identical} (i.e., the lines in Fig.~\ref{fig:vs-T/T} cross). This implies that two networks with a different number of nodes can produce the same amount of entropy (at the same temperature). This is somewhat surprising, as one would expect that a larger system automatically has a higher EP. 

The EP diverges if one of the temperatures goes to zero, corresponding to the cases of vanishing or infinite measurement errors. This is in accordance with the result from~\cite{Strasberg2013}, whose limit of ``precise and infinitely fast'' control corresponds in our time-continuous model to the limit $\mathcal{T'}\to 0$.
We also note that when we interpret $X_0$ as the position of an active swimmer, $\mathcal{T}_0\to 0$ corresponds to the limit of zero translational (``passive'') noise (and pure ``active'' fluctuations). The divergence of EP in this limit was noted in~\cite{pietzonka2017entropy} for a similar active swimmer model.
%
%
%
%
%
\subsection{Limit of discrete delay}
In realistic setups, a controller has a finite memory capacity, i.e., the amount of stored information is limited. 
This implies, on the one hand, that $\mathcal{T'}$ is finite (finite precision), and on the other hand, that the number of memory cells $n$ should be finite. The resulting memory kernel has a peak around $\tau$ with non-zero width (distributed delay).
(Stepping away from the concrete interpretation of the feedback controller with memory cells, distributed delay is 
also the more realistic scenario for many delayed processes, e.g., in biological contexts~\cite{Rateitschak2007,Longtin2010}.)

However, in theoretical studies, the delay is often assumed to be discrete, i.e., infinitely sharp.
In our approach, this corresponds to the limit $n\to \infty$, where $K(T)\to \delta(T-\tau)$. 
In this limit, we find a surprising result for the colored noise. As shown in the Appendix~\ref{sec:limitNoise}, the noise correlations completely \textit{vanish}
\begin{align}
\lim_{n\to \infty} C_\xi = 0
\end{align}
(irrespective of the value of $\mathcal{T}'$).
Hence, a delayed equation with {white} noise is recovered from~(\ref{eq:LE-X0}), that is,
\begin{align}\label{eq:discrete_delay}
\dot{X}_0(t) = a_{00}X_0(t) +k \, X_0(t-\tau) + \xi_0(t).
\end{align}
This implies that the heat flow~(\ref{eq:Q0_general}) approaches the value of the heat flow for discrete delay and white noise, which we have intensely studied in~\cite{Loos2019}, for both, linear and nonlinear systems. The corresponding values (the formula is given in the Appendix~\ref{sec:Q0limitn}) are plotted as dashed lines in Fig.~\ref{fig:vs-n}. 
Indeed, the analytical results for finite $n$ seem to converge to this value. The observed convergence is confirmed by Brownian dynamics simulations for much larger $n>10$ (not shown here).

Remarkably, we generally find that $\dot{Q}_0$ approaches the limit fastest for isothermal conditions $\mathcal{T}_0=\mathcal{T'}$. Thus, the details of the memory kernel (which depend on $n$) seem to have a larger impact, if the temperatures 
are inhomogeneously distributed over the network.

\blue{For $\mathcal{T'} \sim n$, a white noise remains at temperature $\mathcal{T}+(k\tau/\gamma')\mathcal{T'}$ \green{Die Rechnung wuerde ich ins PRE machen, (hier hoechstens erwaehnen). Im PRL koennte man sich auf den Fall beschraenken, dass alle Temperaturen gleich sind}\\}
%
As we have discussed above, the total entropy production increases quadratically which $n$, which implies a \textit{divergent} EP in the limit $n \to \infty$.
%
Taking the perspective that the network models a memory device, where each memory cell can contribute to dissipation, a divergent EP is indeed somewhat expected due to the infinite system size. Also from an information-theoretical viewpoint, this result is in fact not too surprising. 
To realize~(\ref{eq:discrete_delay}), a trajectory of length $\tau$ needs to be memorized. However, the information content of a Brownian trajectory is infinite (in fact, even if it is arbitrarily short) since it is governed by \textit{white noise}, which yields infinitely many jumps within arbitrarily short time intervals. This infinite amount of information results in a divergent EP.
%

Thermodynamic notions of systems with discrete delay were recently also discussed by Rosinberg and Munakata~\cite{Munakata2014,Rosinberg2015,Rosinberg2017}. 
They introduced a framework to define meaningful irreversibility measures based on $X_0$ alone, for the case of underdamped motion. Since we here embed this delay problem into the standard framework of thermodynamics (and do not use a specifically adapted formalism), we find a diverging EP owing to the infinite-dimensional character of this case (which is apparent from the infinite network size, as well as the infinite-dimensional nature of the delta-distribution kernel).
\section{Irreversibility $\&$ Coarse-graining}\label{sec:Coarse-graining}
In this last section, we elucidate the relationship between our network model and the representation via $X_0$ alone from a more fundamental perspective in the context of \textit{irreversibility}.
Considering $X_0$ as the only observable (non-Markovian) process~(\ref{eq:LE-X0}), is, in fact, the viewpoint followed in several recent 
publications, see e.g. ~\cite{caprini2019entropy,dabelow2019irreversibility,Munakata2014,Rosinberg2015,Rosinberg2017,shankar2018hidden}. 
Indeed, there are situations of particular interest, where $X_0$ is the only experimentally accessible quantity. Then, a natural ``irreversibility measure'' which can directly be calculated from the observed trajectories, would be~\cite{dabelow2019irreversibility}
\begin{align}\label{eq:S0}
\frac{\Delta s^\mathrm{0}}{k_\mathrm{B}} := \ln \frac{\mathcal{P}[\mathbf{X}_0 ] }{\mathcal{ \hat{P} }[ \hat{\mathbf{X}}_0 ]}.
\end{align}
With this picture in mind, one may question the meaning of the formula for the total EP  $\Delta s_\mathrm{tot}$, which explicitly involves $X_{j>0}$. However, our framework also yields a statement about $\Delta s^\mathrm{0}$.
First, we notice that $\Delta s^\mathrm{0}$ corresponds to one part of the total EP~(\ref{eq:Stot_general}), i.e., 
\begin{align}\label{eq:stot_parts}
\frac{\Delta s_\mathrm{tot}}{k_\mathrm{B}}= \frac{\Delta s^\mathrm{0}}{k_\mathrm{B}} 
+ \ln \frac{\mathcal{P}[\mathbf{X}_\mathrm{1},..,\mathbf{X}_{n} |\mathbf{X}_0]}{\mathcal{\hat{P}}[ \hat{\mathbf{X}}_\mathrm{1},.., \hat{\mathbf{X}}_{n}|\hat{\mathbf{X}}_0 ]}  .
\end{align}

Further, by exploiting our Markovian representation of the process, we can show that the fluctuations of $s^\mathrm{0} $ fulfill an integral fluctuation theorem, as do those of $\Delta s_\mathrm{tot}$ (see Fig.~\ref{fig:Distributions} and Appendix~\ref{sec:IFT}), 
\begin{align}\label{EQ:Marginalized_IFT}
\left\langle e^{ -\frac{\Delta s^0}{k_\mathrm{B}} }\right\rangle
=&
\int ..\int   
\frac{\mathcal{\hat{P}}[ \hat{\mathbf{X}}_0 ]}{\mathcal{P}[ \mathbf{X}_0 ]}
\,
\mathcal{P}[ \mathbf{\underline{X}} ]
\,
\mathfrak{D}\mathbf{\underline{X}} 
\nn
= &
\int \!
{\mathcal{\hat{P}}[ \hat{\mathbf{X}}_0 ]}
\underbrace{
\int \!
\mathcal{P}[ \mathbf{X}_1,..,\mathbf{X}_n| \mathbf{X}_0 ]
\,
\mathfrak{D}\mathbf{X}_1.. \mathfrak{D}\mathbf{X}_n
}_{=1}
\,
\mathfrak{D}\mathbf{X}_0 
\nn
= &
\int 
{\mathcal{\hat{P}}[ \hat{\mathbf{X}}_0 ]}
\,
\mathfrak{D}\mathbf{X}_0
= 1.
\end{align}
This implies $ \Delta s^\mathrm{0}  \geq 0 $. The same can be shown for the other contribution in~(\ref{eq:stot_parts}) (see in the Appendix~\ref{sec:IFT}), yielding the inequality
\begin{align}\label{eq:inequality_S0-Stot}
 \Delta S_\mathrm{tot}   \geq  \Delta S^\mathrm{0}  \geq 0  .
\end{align}
Thus, the irreversibility of $X_0$ alone underestimates the irreversibility of the entire system, consistent with what one would expect~\cite{Esposito2012}. Furthermore,~(\ref{eq:inequality_S0-Stot}) states that the mean total EP considered in this paper yields an upper bound to the average of the irreversibility measure $\Delta s^\mathrm{0}$.

Importantly, the fluctuation theorems can only be shown, when the ensemble average $\langle ... \rangle = \iint ...\mathcal{P}[ \mathbf{\underline{X}} ] \mathfrak{D}\mathbf{\underline{X}} $ is evaluated using the \textit{full} path probability $\mathcal{P}[ \mathbf{\underline{X}} ]$, while averages with respect to marginalized measures do not yield the correct relations.

By considering $\Delta s^\mathrm{0}$, we have addressed entropy-like quantities based on marginalized path probabilities. An obvious alternative to get rid of the additional d.o.f. $X_{j>0}$, would be to study instead the marginalized EP: $\iint \Delta  s_\mathrm{tot}[\mathbf{\underline{X}}]\, \mathfrak{D}\mathbf{X}_1...\mathfrak{D}\mathbf{X}_n $. We stress that this quantity is different from $\Delta s^\mathrm{0}$ [as can be seen from~(\ref{eq:stot_parts})] and there is no direct relation between both. The relevance of such a marginalized EP is, in fact, questionable, as it does \textit{not} solely rely on measurable quantities (as opposed to $\Delta s^\mathrm{0}$), neither does it have a direct link to the well-understood framework of statistical mechanics. In contrast, the total EP $\Delta s_\mathrm{tot}$ is the central quantity in statistical mechanics with a direct connection to the energy flows, the first and second law of thermodynamics, and the underlying microscopic laws of motion~\cite{Wittkowski2013}.

\section{Conclusion \label{sec:Conclusion}}
\subsection{Summary}
This paper addresses the long-standing problem of finding a thermodynamic description for non-Markovian, active systems with memory and feedback. As opposed to earlier studies focusing on a one-variable description, we here discuss an alternative approach which explicitly takes into account those parts of the environment that generate the memory. 
By doing so, we acknowledge that in the real world, memory is indeed due to the interaction with other subsystems. 
{We find that NR coupling between d.o.f. plays a key role.} To establish the thermodynamic implications,
we have analytically calculated central quantities such as non-zero energy input, which we identify with activity, and information flow, 
the main characteristic of sensing.
 
We have shown that feedback can be viewed as a combination of measurement and active force. Moreover, when the measurement is the dominating operation (i.e., the controller ``knows'' more about the sensed system than vice versa, indicated by a positive information flow out of the system), some major goals of feedback control can be achieved, including thermal fluctuation suppression (i.e., feedback cooling), and a reversed heat flow (i.e., energy extraction of the heat bath).
We demonstrate that all these features can already be described by a minimal, linear model consisting of only two d.o.f.. Activity and feedback emerge as two phenomena describable by the same concept, that is, NR coupling. 
 
Further, we have shown that NR interactions are an essential ingredient to generate intricate, non-monotonic memory kernels. A unidirectionally coupled ring network yields a kernel that is peaked around a delay time, and can be used to model a Brownian particle subject to an optical tweezers feedback control. %
The special case of discrete delay with white noise emerges as the limit of an infinitely long ring in our approach. We find that the heat exchange of the system of interest, $X_0$, approaches a well-known result from the literature~\cite{Loos2019}. On the contrary, the total entropy production diverges, suggesting that the cost of storing an infinite amount of information is unbounded.

Taken together, from our analysis emerges a unifying perspective on activity, feedback and memory. This underlines the importance of a thermodynamic and information-theoretic consideration.

One important question we have not yet addressed, is the uniqueness of a Markovian representation, i.e., is it possible to find the exact same memory in another network of different size, coupling topology, or with another type of dynamics? We think that the representation is indeed not unique. In fact, exploring more complex network architectures (which potentially allow to represent the same memory via less d.o.f.), is an interesting question and subject of future work. However, by considering linearly coupled, overdamped nodes with linear dynamics, we have, in fact, studied the most simple possibility. 
The Gamma-distributed delays which are automatically created by this architecture have indeed many applications. For example, in biological contexts they appear in gene regulatory networks~\cite{lai2016understanding,Josic2011,Rateitschak2007,Friedman2006}, or in the dynamics of virus spreading between cells~\cite{Mittler1998}. 
We note that for a general functional form of the memory kernel, a finite-dimensional representation is not guaranteed. Combining Gamma-distributed kernels, we obtain models for a very general type of memory. In fact, any kernel that eventually decays exponentially can be approximated in this way. We find that a memory kernel with $N$ extrema can be represented by $N$ NR coupled subsystems. Reciprocally coupled d.o.f., on the other hand, yield only monotonically decaying memory kernels.

At the end of this paper, we have discussed situations where only some of the relevant d.o.f., e.g., only $X_0$, are observable. We shall address the question: ``What is then the purpose of a framework involving all d.o.f.?''. First of all, we have shown that it indeed also yields statements about entropy-like measures appropriate for such situations. More importantly, some key quantities, like the heat flow of $X_0$, 
are indeed not affected by the choice of description, i.e., whether one chooses to employ a non-Markovian Langevin description, or a Markovian embedding approach. 

%
%
\subsection{Outlook and further perspectives}

We close this article with some comments putting our approach in a broader perspective.

There are important links to computer science. Our approach might be useful to incorporate the thermodynamic cost of memory in the theoretical description of computation in living~\cite{kempes2017thermodynamic} and artificial systems~\cite{wolpert1992memory}. 
Further, the ring network architecture studied here is very similar to the reservoir computers investigated in~\cite{larger2017high,li2018deep}. A reservoir computer of this type may be experimentally realized by a laser network~\cite{rohm2019reservoir,rohm2018multiplexed}, or by coupled RC circuits~\cite{kish2012electrical,snider2011minimum}. Another link to machine learning is the similarity between the ring network and recurrent neural networks~\cite{Xu2004}, used for example for reinforcement learning. In these contexts, the connection between NR coupling and information flow discussed here might be of particular importance. Noteworthy, the topology of the unidirectionally coupled ring considered here also resembles the architecture of a Brownian clock~\cite{barato2016cost}, which, in contrast, has discrete dynamics.

A conceptionally interesting aspect regarding the way of modeling memory employed in this work, is that it can be considered as a ``discretization of the past''~\cite{Loos2019b}. This is a delicate point, as the dynamics itself is time-\textit{continuous}. As opposed to that, a discretization of the entire dynamics plus past trajectory, was proposed in~\cite{kwon2017information}. Surprisingly, the approach taken here yields a simpler thermodynamic framework, despite the apparent contradiction of its discrete--continuous nature.

We have 
pointed out fundamental connections between the thermodynamic properties of individual active and feedback-controlled systems with memory. 
Given these connections, one shall also expect resemblance of their collective behavior. Indeed, clustering and pattern formation, whose occurrence is well-known in active systems, particularly in microswimmer suspensions~\cite{ramaswamy2010mechanics,ramaswamy2017active,nagai2015collective,kaiser2017flocking}, were recently also reported in collective feedback-controlled systems~\cite{tarama2019traveling,Mijalkov2016,Khadka2018,Holubec2019}. Thereby, the delay is found to play a crucial role.
We also note that in non-linear dynamics and network science, studying the interplay between symmetry-broken coupling and collective behavior is already a well-established research field~\cite{Loos2016}. Indeed, the existence of chimera states, a special type of clustering, was linked to symmetry-broken coupling~\cite{Premalatha2015}, and shown to persist in the presence of discrete delay~\cite{Zakharova2016} and Gamma-distributed memory~\cite{kyrychko2013amplitude}.
It will be interesting to see how these collective phenomena can be treated by means of thermodynamics (whose relevance for collective behavior of passive, equilibrium systems is undeniable). Understanding the thermodynamic properties of individual constituents is a first step in this direction.

Also in quantum systems, researchers have recently started to analyze, on the one hand, feedback with memory~\cite{strasberg2013thermodynamics,Carmele2013}, and, on the other hand, non-reciprocal interactions~\cite{Fang2017,Metelmann2015}. 
Here, we unify these two concepts for classical (stochastic) models. It would be very interesting to generalize this perspective to the quantum world, where entropy, information and energy play an analogous role, while measurement and stochasticity have different notions.

An interesting way to think about the approach discussed in this paper is that it points out a connection between time-reversal symmetry breaking and symmetry-broken coupling topology. The entropy production, which is the \textit{thermodynamic arrow of time}, directly measures the breaking of time-reversal symmetry, i.e., how sure can we be that a system's evolution is running from past to future, and not vice versa. In a way, memory breaks that symmetry in a fundamentally different way than, e.g., an external field, because memorizing is \textit{per se } only one-way in time.

\begin{acknowledgments}
This work was funded by the Deutsche Forschungsgemeinschaft (DFG, German Research Foundation) - Projektnummer 163436311 - SFB 910. 
\end{acknowledgments}
\appendix
\section{\uppercase{Memory kernel and noise correlations}}\label{sec:deriveKandNu}
We here derive the memory kernel and colored noise in the model~(\ref{K_noise_II}), i.e., a unidirectional ring network of arbitrary length $n+1$.
To this end, we solve the equations for $X_{j\in \{1,2,...,n\}} $ in frequency space, making use of their linearity (we want to emphasize that their linearity is irrespective of the question whether the equation of $X_0$ is linear). 
First, we apply the Laplace transformation
 $\mathcal{L}[X_j(t)](s)=\int_{0}^{\infty} X_j(t)e^{-st}\mathrm{d}s$ 
 and obtain for each $j\in \{1,2,...,n\}$
\begin{align}\label{eq:Laplace_Xj_all}
 s {\hat{X}_j}(s)= &({n}/{\tau}) \left[\hat{X}_{j-1}(s)- {\hat{X}_j}(s)\right]+ \,\hat{\xi}_j(s)/\gamma' \nn
\Rightarrow {\hat{X}_j}(s)=& \frac{({n}/{\tau})\hat{X}_{j-1}(s)  + \hat{\xi}_j(s)/\gamma'}{ s+({n}/{\tau})}
,
\end{align}
and by iteratively substituting the solutions (\ref{eq:Laplace_Xj_all}),
\begin{align}\label{eq:Laplace_Xn}
{\hat{X}_n}=& \frac{ ({n}/{\tau})^n }{[ s+({n}/{\tau})]^n}\hat{X}_{0}+\sum_{j=1}^{n}\frac{ ({n}/{\tau})^{j-1} \,\hat \xi_{n-j+1}/\gamma' }{[ s+({n}/{\tau})]^{j}}
.
\end{align}
Now we transform back to the real space via inverse Laplace transformation. In~(\ref{eq:Laplace_Xn}) we identify the Laplace-transform of the Gamma-distribution
${\displaystyle \mathcal{L}\left[K_{j}(t)\right]( s)={\frac {(n/\tau)^{j}}{[s+(n/\tau)]^{j}}}} $ with the Gamma-distributed kernels $K_{j}(t)={\frac {n^{j}}{\tau^{j} (j-1)!}}\,t^{j-1}e^{-n t /\tau}$. We further make use of the convolution theorem as well as the linearity of the Laplace transformation, and find
\begin{align}\label{eq:RealSpace_Xn}
{X}_n(t)
%
=& \int_{0}^{t} K_{n}(t-t') {X}_{0}(t') \,\mathrm{d}t' + \nu_n(t)
\end{align}
with the Gaussian colored noise
\begin{align}
\nu_n(t)=&\sum_{j=1}^{n}\int_{0}^{t} \frac{\tau}{n} K_{j}(t-t')  \, \xi_{n-j+1}(t') \,\mathrm{d}t'.  
\end{align}
Due to the coupling between $X_n$ and $X_0$ in~(\ref{eq:Network}), this yields the memory kernel $K =  K_{n}$ and the colored noise $\nu = k \nu_n $ in the dynamical Eq.~(\ref{eq:LE-X0}) of $X_0$.
The noise correlations $C_{\nu}(\Delta t)=\langle \nu(t)\nu(t+\Delta t)\rangle$ can be calculated exactly for an arbitrary $\Delta t>0$, as we will show in the following. First, we integrate out the $\delta$-correlated noise terms
, yielding 
\begin{align}\label{eq:colorednoise_1}
\frac{C_{\nu}(\Delta t)}{\psi}=&
\sum_{j=1}^{n}\int_{0}^{t} 
 K_{j}(t-t')K_{j}(t-t'+\Delta t)\,\mathrm{d}t'
\nn =& \sum_{j=1}^{n} 
 \frac {(n/\tau)^{2j}}{ (j-1)!^2} 
 \int_{0}^{t}  (t'^2+t' \Delta t )^{j-1}e^{-n (\Delta t+2t')/\tau}\,\mathrm{d}t',
\nn
 =&e^{-n \Delta t /\tau}\sum_{j=1}^{n}
 \frac {(n/\tau)2^{1-2j }}{(j-1)!^2} ~{(*)},
\end{align}
with $\psi ={2  k_\mathrm{B}\mathcal{T}' (\tau/n)^2 k^2 / \gamma'}$ and
\begin{align}
{(*)}= \int_{0}^{2tn/\tau} \left(u^2+u\frac{2n}{\tau} \Delta t\right)^{j-1}e^{-u}\,\mathrm{d}u .
\end{align} 
In the last step we have performed the substitution $u=2t' n/\tau$ and canceled several factors. The integral ${(*)}$ can be simplified by using the binomial theorem,
which yields
\begin{align}
{(*)}=&
\sum_{l=0}^{j-1} \binom{j-1}{l} \int_{0}^{2tn/\tau} u^{2(j-1-l)} \left(u\frac{2n\Delta t}{\tau} \right)^l
 e^{-u}\,\mathrm{d}u
 \nn&
\sum_{l=0}^{j-1} \frac{\left(2n\Delta t\right)^l (j-1)!}{\tau^l l!(j-l-1)!} \int_{0}^{2tn/\tau} u^{2j-l-2}
 e^{-u}\,\mathrm{d}u.
\end{align} 
We take the limit $t \to \infty$, since we focus on steady states and are mainly interested in the noise correlations after the initial condition effects have decayed. Then, we perform the integration using
$
\int_{0}^{\infty} x^p e^{-x}\mathrm{d}x =p!,
$ 
which yields
\begin{align}\label{eq:colorednoise_*}
{(*)}=
\sum_{l=0}^{j-1} \left(\frac{2n\Delta t}{\tau} \right)^l\frac{(j-1)!(2j-l-2)!}{l!(j-l-1)!}.
\end{align}
(Transient dynamics could be treated similarly by using the incomplete Gamma function.)
Combining (\ref{eq:colorednoise_1}, \ref{eq:colorednoise_*}) yields 
the noise correlation given in~(\ref{K_noise_II}). 
%
%
%
%
%
%
\section{\uppercase{FDR for unidirectional ring of arbitrary length}}\label{sec:FDT_modelII}
In this Appendix, we generally consider the existence of the FDR for a network of type (\ref{K_noise_II}) with arbitrary length $n$. The FDR is discussed in the main text in Sec.~\ref{sec:FDR}. 
We need to show the equivalence of the following two terms
\begin{align}\label{eq:CheckFDR-1} 
&k_\mathrm{B} \mathcal{T}_0{\gamma(\Delta t)}= k_\mathrm{B} \mathcal{T}_0 \left[  2 \gamma_0 \,\delta(\Delta t) +    \frac{ k  \Gamma\left( n, \frac{n  \Delta t}{\tau} \right)} {(n-1)!}  \right]
\end{align}
and
\begin{align}\label{eq:CheckFDR-2}
&\langle \mu(t) \mu(t+\Delta t) \rangle = \, {k_\mathrm{B} \mathcal{T}_0}  \,2 \gamma_0  \delta(\Delta t) \nonumber \\
& +   \frac{k_\mathrm{B} k^2}{\gamma'} \sum_{p=0}^{n-1} \sum_{l=0}^{p} \mathcal{T}'   \frac{2^{l-2p}(2p-l)!}{p!l!(p-l)!} \frac{e^{-n \Delta t /\tau} \Delta t^l}{(\tau/n)^{l-1}} . %
\end{align}
The first summands on the right hand side of~(\ref{eq:CheckFDR-1}) and~(\ref{eq:CheckFDR-2}) are already equivalent.
Further, the other term in~(\ref{eq:CheckFDR-1}) an be rewritten  by partially integrating the incomplete Gamma function as
\begin{align}
\Gamma\left( n, {n  \Delta t}/{\tau} \right)
= &(n-1)!\,    \sum_{p=0}^{n-1} \left(\frac{n\Delta t}{\tau}\right)^{p} \frac{e^{-n\Delta t/\tau}}{p!}  .
\end{align}
Thus, the FDR holds, if at all $\Delta t \geq 0$
\begin{align} \label{EQ:FDT_nec0} 
 \sum_{p=0}^{n-1}   \sum_{l=0}^{p} \frac{\mathcal{T}'}{\mathcal{T}_0}\left[ \frac{k \,2^{l}(2p-l)! }{n\,2^{2p} l!(p-l)!}\right]
  \frac{\tau(n\Delta t)^l e^{-n \Delta t /\tau} }{  \gamma'\, \tau^{l} p!} 
  \stackrel{?}{=} 
 \nn
\sum_{p=0}^{n-1} \frac{ (n\Delta t)^{p} e^{-n\Delta t/\tau}}{\tau^p  p!} .
\end{align}
At $\Delta t =0 $ Eq.~(\ref{EQ:FDT_nec0}) is fulfilled, if and only if
\begin{align}\label{EQ:Cond_1}
\mathcal{O}(\Delta t^0):~~\frac{\gamma'}{\tau } \stackrel{?}{=}   \sum_{p=0}^{n-1}   \frac{\mathcal{T}'}{\mathcal{T}_0}  \frac{k  (2p)!}{n\,2^{2p} (p!)^2}.
\end{align}
This is the zero$^\mathrm{th}$-order condition (in $\Delta t$). For $n=1$, Eq.~(\ref{EQ:Cond_1}) reads $k\tau {\mathcal{T}' }= \gamma'{\mathcal{T}_0 }$ [in agreement with~(\ref{eq:Fulfill-FDR})], and is a \textit{sufficient} condition. It can only be fulfilled for $k \geq 0$. Further, for $n>1$, it yields a \textit{necessary}, but not sufficient condition due to higher order contributions in Eq.~(\ref{EQ:FDT_nec0}). 
For example, the first order in $\Delta t$, which is relevant for all $n >1$, yields the additional condition
\begin{align} 
\mathcal{O}(\Delta t^1):~~ \frac{\gamma' }{\tau   } (n\Delta t) 
  \stackrel{?}{=} 
  \frac{\mathcal{T}'}{\mathcal{T}_0}  \frac{k  }{2\,n } 
  (n\Delta t) .
\end{align}
Combining this with~(\ref{EQ:Cond_1}) gives
\begin{align} \label{EQ:FDT_nec} 
\sum_{p=0}^{n-1}   \frac{\mathcal{T}' (n\Delta t)}{\mathcal{T}_0}  \frac{ 2 (2p)!}{\,2^{2p} (p!)^2}
  \stackrel{?}{=} 
  \frac{\mathcal{T}'  (n\Delta t)}{\mathcal{T}_0}
  ,
\end{align}
%
which can only be fulfilled if $\mathcal{T}'=0$. However, at $\mathcal{T}'=0$ condition~(\ref{EQ:Cond_1}) is violated. Hence, the FDR is broken for all choices $\mathcal{T}'$ when $n >1 $. 

We have, in fact, also investigated a generalization of (\ref{K_noise_II}), where all temperatures are allowed to be different, i.e., $X_j$ has temperature $\mathcal{T}_j$ (generalizing the derivations presented in Appendix~\ref{sec:deriveKandNu} is straightforward). Even in this case, the FDR cannot be fulfilled if $n>1$.
\section{\uppercase{Fluctuation theorems}\label{sec:IFT}}
We consider the fluctuation theorems discussed in Sec.~\ref{sec:Coarse-graining} in the main text. We recall that $\underline{X}=(X_0 ,.., X_{n})^T$ denote vectors, $\mathbf{\underline{X}}=\{\underline{X}(t)\}_{t_i}^{t_f}$ are trajectories, while (joint) path probabilities involving all d.o.f. are denoted $\mathcal{P}[ \mathbf{\underline{X}} ]$.
%
First, we show the integral fluctuation theorem for the total EP using the joint path probabilities
\begin{align}\label{eq:IFT}
\left\langle 
e^{ -\Delta s_\mathrm{tot}/k_\mathrm{B}} 
\right\rangle
= &
\iint e^{-\Delta s_\mathrm{tot}[ \mathbf{\underline{X}} ]}
\mathcal{P}[ \mathbf{\underline{X}} ]
\,
\mathfrak{D}[ \mathbf{\underline{X}} ]
\nn= &
\iint
e^{\ln \frac{\mathcal{\hat{P}}[ \hat{\mathbf{\underline{X}}} ]}{\mathcal{P}[ \mathbf{\underline{X}} ]}}
\mathcal{P}[ \mathbf{\underline{X}} ]
\,
\mathfrak{D}[ \mathbf{\underline{X}} ]
\nn= &
\iint
\mathcal{\hat{P}}[ \hat{\mathbf{\underline{X}}} ]
\,\mathfrak{D}[ \mathbf{\underline{X}} ] = 1.
\end{align}

Further, we note that relations of type $\langle  e^{-s} \rangle = 1$ generally imply $\langle  s \rangle\geq 0$, which follows from $ \ln(x) \leq  x-1$. This argument can be applied when the ensemble average is expressed as sum over $N \to \infty$ realization of the random process: 
$\langle -s \rangle = \sum_{i=1}^{N} \frac{-s_i}{N}  \leq \sum_{i} \frac{e^{-s_i}-1}{N}= \langle  e^{-s} \rangle -1=0
$. 

Now, we use the definition of conditional path probabilities, i.e., 
$\mathcal{P}[  \mathbf{X}_0;\mathbf{\underline{X}}_\mathrm{c} ]= \mathcal{P}[ \mathbf{X}_0|\mathbf{\underline{X}}_\mathrm{c} ]\mathcal{P}[ \mathbf{X}_0 ]$, with $\mathbf{\underline{X}}_\mathrm{c}=\{\mathbf{{X}}_1,...,\mathbf{{X}}_n\}$, to split up the total EP
\begin{align}
\frac{\Delta s_\mathrm{tot}[\mathbf{\underline{X}}]}{k_\mathrm{B}}
=
\ln \frac{\mathcal{P}[ \mathbf{\mathbf{\underline{X}}} ]}{\mathcal{\hat{P}}[ \hat{\mathbf{\underline{X}}} ]}
=
\frac{s^0[ \mathbf{X}_0]}{k_\mathrm{B}}
+
\underbrace{\ln \frac{\mathcal{P}[ \mathbf{X}_\mathrm{c}| \mathbf{X}_0 ]}{\mathcal{\hat{P}}[ \hat{\mathbf{\underline{X}}}_\mathrm{c}| \hat{\mathbf{X}}_0 ]}}_{ =s^\mathrm{c|0}[ \mathbf{X}_0;\mathbf{\underline{X}}_\mathrm{c}   ]/k_\mathrm{B}}
.
\end{align}
[equivalently to~(\ref{eq:stot_parts})].
The entropy-like quantity $\Delta s^\mathrm{c|0}$ as well fulfills an integral fluctuation theorem 
 \begin{align}
\left\langle e^{-\Delta S ^\mathrm{c|0}/k_\mathrm{B}}\right\rangle
= &
\iint   
\frac{\mathcal{\hat{P}}[\hat{\mathbf{\underline{X}}}_\mathrm{c}| \hat{\mathbf{X}}_0] }{\mathcal{P}[\mathbf{\underline{X}}_\mathrm{c}| \mathbf{X}_0  ]}
\mathcal{P}[ \mathbf{\underline{X}} ]
\,\mathfrak{D}\mathbf{\underline{X}} 
\nn
= &
\int
\mathcal{P}[ \mathbf{X}_0 ]
\left[
\int 
\mathcal{\hat{P}}[\hat{\mathbf{\underline{X}}}_\mathrm{c}| \hat{\mathbf{X}}_0 ]
\,\mathfrak{D}\mathbf{\underline{X}}_\mathrm{c} 
\right]
\,
\mathfrak{D}\mathbf{X}_0
\nn= &
\int 
{\mathcal{P}[ \mathbf{X}_0 ]}
\,
\mathfrak{D}\mathbf{X}_0
 = 1.
\end{align}
[The same holds for $\Delta s^\mathrm{0}$, see Eq.~(\ref{EQ:Marginalized_IFT}).]
%
%
%
\section{\uppercase{``Mild cases''}}\label{sec:hiddenTemp}
 %
We here show that for the specific case $a_{01}a_{01}\!>\!0$, the NR coupling can potentially be traced back to a hidden temperature gradient and represents hence a ``mild'' form of intrinsic non-equilibrium, as noted in Sec.~\ref{sec:Energy}.
Consider the NR network of $X_0$ and $X_1$ with $a_{01}a_{01} \neq 0$ 
\begin{align} 
\begin{cases}\label{eq:ASymmetric-network}
\gamma_0 \dot{X}_0 &=a_{00}X_0 +a_{01}X_1 + \xi_0\\
\gamma_1 \dot{X}_1 &=a_{10}X_0 +a_{11}X_1 + \xi_1. 
\end{cases}
\end{align} 
Via a scaling, (\ref{eq:ASymmetric-network}) can be mapped onto another network, in particular, by introducing $\widetilde{X}_0=\sqrt{| a_{10} |}\,X_0$, $\widetilde{X}_1=\sqrt{| a_{01} |}\,X_1$, and $\widetilde{\mathcal{T}}_0 =|a_{10}|\mathcal{T}_0$, $\widetilde{\mathcal{T}}_1 =|a_{01}|\mathcal{T}_1$. (We note that scaling of positional d.o.f. shall indeed be accompanied with scaling of the temperatures due to the connection between temperatures and the time-derivative of the positions, e.g., think of  the equipartition theorem). In this way, we find
\begin{align}
\begin{cases}\label{eq:Symmetric-network}
\dot{\widetilde{X}}_0 &=  a_{00} \widetilde{X}_0 + \text{sgn}(a_{01})\sqrt{a_{01}a_{10} } \tilde{X}_1 + \widetilde{ \xi}_0\\
\dot{\widetilde{X}}_1 &= \text{sgn}(a_{10}) \sqrt{ a_{10}a_{01}  }  \widetilde{X}_0 +  a_{11} \tilde{X}_1 + \widetilde{\xi}_1,
\end{cases}
\end{align}
with $\langle \widetilde{ \xi}_i(t)\widetilde{ \xi}_j(t) \rangle =2 k_\mathrm{B}\widetilde{\mathcal{T}}_j\gamma_j \delta_{ij}\delta(t-t')$.
If $a_{01}a_{01}\!>\!0$, this system has \textit{reciprocal} coupling. Further, even if $\mathcal{T}_0=\mathcal{T}_1$, it involves a temperature gradient. This symmetric network~(\ref{eq:Symmetric-network}) could, for example, model the angles of two vanes in different heat baths, coupled by a torsion spring~\cite{Sekimoto2010}. 
As well-known, such a reciprocally coupled system 
equilibrates if, and only if, 
$\widetilde{\mathcal{T}}_1 = \widetilde{\mathcal{T}}_0 $ $\Leftrightarrow |a_{01}|{\mathcal{T}}_1= |a_{10}|\mathcal{T}_0 $. This is identical to the equilibrium-condition~(\ref{eq:Fulfill-FDR}).

Now we turn to the impact of this scaling on the thermodynamic quantities. For the energy flows, we find the relations
\begin{align}
{\delta \widetilde{w}}_\mathrm{0}  &=  \text{sgn}(a_{01})\sqrt{a_{01}a_{10} }\, \widetilde{X}_1  \circ \mathrm{d}{\tilde{X}}_0 
\nn
&=   a_{01} |a_{10}|\,X_1  \circ \,\mathrm{d}{X}_0 =   |a_{10}|\,{\delta   w }_\mathrm{0}  ,
\nn
{\delta \widetilde{q}}_\mathrm{0}  &=   (\gamma_0 \dot{\widetilde{X}}_0-\widetilde{\xi}_0 ) \circ \mathrm{d}{\tilde{X}}_0 =  |a_{10}|\,{\delta q }_\mathrm{0}  . 
\end{align} 
Thus, the energy flows to and out of $X_0$ are both scaled with $|a_{10}|$. Likewise, $\delta \tilde{w}_1 =  |a_{01}|\, \delta w _\mathrm{1}$ and $\delta \tilde{q}_1 =  |a_{01}|\, \delta q _\mathrm{1}$. This further means
\begin{align}
\Delta {\tilde{s}}_\mathrm{tot} =& \frac{  {\delta \tilde{q}}_\mathrm{0} } {|a_{10}|\mathcal{T}_0}
+ \frac{{\delta \tilde{q}}_\mathrm{1} } {|a_{01}| \mathcal{T}_1}=\frac{  {\delta {q}}_\mathrm{0} } {\mathcal{T}_0}
+ \frac{{\delta {q}}_\mathrm{1} } { \mathcal{T}_1}=\Delta {s}_\mathrm{tot},
\end{align}
i.e., the EP in the scaled model is \textit{identical} to the EP in the original model, implying that equilibration of \textit{both} models is expected, if~(\ref{eq:Fulfill-FDR}) holds. 

We conclude that the two ``driving mechanisms'', that is, NR coupling with $a_{01}a_{10}>0$, or a temperature gradient, can formally not be distinguished on the level of EP. It should be emphasized that a scaling as employed here cannot be found if $a_{01}a_{10}\leq 0$. Thus, the NR coupling considered here is the more general case.

%
%
%
%
%
\section{\uppercase{Mutual information}}\label{sec:AppInfo}
Here we discuss the relation between the information flow considered in Sec.~\ref{sec:Info}, and the mutual information.
We apply basic properties of the logarithm and the natural boundary conditions to show that the information flow~(\ref{eq:Iflow}) (note the different sign convention of ${\dot{I}}_{\to j} $ as compared to~\cite{horowitz2014second}) can be expressed as
\begin{align}
{\dot{I}}_{\to j} 
= &\iint \! \ln \frac{\rho_{1}(x_j)}{  \rho_{n+1}(\underline{x})}  \partial_{x_j} J_j  \,\mathrm{d}\underline{x}
\nn
= & \iint \! \ln \frac{ \rho_{1}(x_j) \rho_{1} (x_{i\neq j}) }{  \rho_{n+1}(\underline{x})}  \partial_{x_j} J_j  \,\mathrm{d}\underline{x} 
\nn
&
-\iint \! \ln \rho_{1}(x_{i }) \underbrace{\left[  J_j \right]_{-\infty}^{\infty}}_{\to 0 } \,\mathrm{d}\underline{x}_{\neq j} 
\nn
=&... =
  \int \! \ln \frac{ \rho_{1}(x_0)\rho_{1}(x_1)..\rho_{1}(x_n) }{  \rho_{n+1}(\underline{x})} \partial_{x_j} J_j  \,\mathrm{d}\underline{x}
  .
 \end{align}
Now, we show the connection to the mutual information between all $n+1$ d.o.f. defined in~(\ref{def:mutualInfo}).
Its total derivative can, by utilizing the FPE~(\ref{eq:FPE-extended}), be expressed as
\begin{align}
\mathcal{\dot{I}} = 
& \int \!  \underbrace{\partial_t \rho_{n+1}(\underline{x})}_{  =-\sum_j \partial_{x_j}J_j  }   \ln \frac{  \rho_{n+1}(\underline{x})}{ \rho_{1}(x_0)...\rho_{1}(x_n) }  \,\mathrm{d}\underline{x} 
\nn
&+ \int \!  \rho_{n+1}  \left[  -\frac{  \partial_t  \rho_{n+1}}{ \rho_{n+1}} - \frac{  \partial_t (\rho_{1}...\rho_{1})}{ \rho_{1}...\rho_{1}}   \right] \mathrm{d}\underline{x}  
\nn
  =&
\sum_{j=0}^n\int \partial_{x_j} J_{j}(\underline{x})   \ln \frac{ \rho_{1}(x_0)...\rho_{1}(x_n) } {  \rho_{n+1}(\underline{x})} \mathrm{d}\underline{x} 
- \int \underbrace{ \partial_t  \rho_{n+1}}_{\to 0} \mathrm{d}\underline{x} 
\nn
&- \int   \frac{\rho_{n+1} }{ \rho_{1}(x_0)...\rho_{1}(x_n)  }  \underbrace{\partial_t[\rho_{1}(x_0)...\rho_{1}(x_n)]}_{\to 0}  \mathrm{d}\underline{x}  
.
\end{align}
Thus, the change of mutual information is given by the sum over all information flows, $\sum_{j=0}^n {\dot{I}}_{\to j}  =\mathcal{\dot{I}}$. (As was shown in~\cite{allahverdyan2009thermodynamic}, the information flow ${\dot{I}}_{\to j}$ is actually the ``time-shifted mutual information'' with the time shift applied to $X_j$.) On the other hand, as can easily be seen from its definition~(\ref{def:mutualInfo}), $\mathcal{{I}}$ is a conserved quantity in steady-states, $\mathcal{\dot{I}}=0$.
Hence, the information flows between all components of the entire network nullify. %

%
%
%
%
%
%
%
%
\subsection*{Information flow for $n=2$}
We recall the general formula for the information flow to $X_j$ given in Eq.~(\ref{eq:generalFormula_Infoflow}). For $n=2$, this results in
\begin{widetext}
\begin{align}\label{eq:info_n=2}
\frac{{\dot{I}}_{\to 0}}{\gamma_0^{-1}} 
&=
\frac{[\langle X_0X_2 \rangle \langle X_1X_2 \rangle - \langle X_0 X_1 \rangle\langle X_2^2 \rangle]\frac{\dot{Q}_0}{\mathcal{T}_0a_{02}}+[\langle X_0 X_1 \rangle\langle X_1X_2 \rangle-\langle X_0 X_2 \rangle\langle X_1^2 \rangle][ a_{00}\langle X_0 X_2\rangle + a_{02} \langle X_2^2 \rangle]  }
{\langle X_0^2 \rangle\langle X_1 X_2 \rangle^2+\langle X_1^2 \rangle\langle X_0 X_2 \rangle^2+\langle X_2^2 \rangle\langle X_0 X_1 \rangle^2 -2\langle X_0 X_1 \rangle\langle X_0 X_2 \rangle\langle X_1 X_2 \rangle-\langle X_0^2 \rangle\langle X_1^2 \rangle\langle X_2^2 \rangle}.
%
%
\end{align}
\end{widetext}
The information flow can be nonzero, despite $\dot{Q}_0 =0$.
%

%
%
\section{\uppercase{Non-Monotonic memory from NR coupling}}\label{sec:MemoryRing}
To illustrate the connection between memory and NR coupling, let us consider a simple $n=2$ model
\begin{align}
\dot{X}_j=-(p+\kappa)X_j +p X_{j+1}+\kappa X_{j-1}+\xi_{j},
\end{align}
with $j \in \{0,1,2\}$. Projecting this set of equations onto $X_0$, by performing a Laplace-transformation and iteratively solving the algebraic equations for $X_0$ [as explained in detail in the Appendix~\ref{sec:deriveKandNu}], we obtain
\begin{align}
 s{\hat{X}}_0 =&  p \left[ \frac{ p^2 }{ (s+p+\kappa)^2- p \kappa  } + \frac{ \kappa (s+p+\kappa)}{ (s+p+\kappa)^2- p \kappa  }  \right]  \hat{X}_0   
 \nonumber \\
 &+\kappa  \left[ \frac{ \kappa^2 }{ (s+p+\kappa)^2- p \kappa  } + \right. \left. \frac{ p (s+p+\kappa)}{ (s+p+\kappa)^2- p \kappa  } \right]  \hat{X}_0  
 \nn
 &
 - \kappa \hat{X}_0 - p \hat{X}_0 +\mathcal{O}(\hat{\xi}_j).
\end{align}
Transforming back to real space yields a non-Markovian process~(\ref{eq:LE-X0})
with the memory kernel~(\ref{eg:kernel_n=3}).
\section{\uppercase{Limit of colored noise}}\label{sec:limitNoise}
Here, we perform the limit of $n\to\infty$ of the noise correlation given in~(\ref{K_noise_II}), which read for $\Delta t\geq 0$
$$
\frac{C_\nu(\Delta t)}{\gamma'  k_\mathrm{B}\mathcal{T}'   k^2}    = \sum_{p=0}^{n-1} \sum_{l=0}^{p} \frac{n^l 2^{l-2p}(2p-l)!}{n \tau^{l-1}p!l!(p-l)!}  e^{-n \Delta t /\tau} \Delta t^l . $$
\blue{
First, we consider the value at $\Delta t=0$, where it simplifies to 
\begin{align}
C_\nu(0) =&\frac{ 1 }{n\tau }  \sum_{m=1}^{n}  \frac{ (2(n-m))!  }{2^{2(n-m)}((n-m)!)^2} 
= \frac{ 1 }{2\tau }  \sum_{p=0}^{n-1}  \frac{ (2p)!  }{2^{2p}(p!)^2} .
\end{align}
In order to further simplify this, we use 
$\Gamma(x+1)=x\Gamma(x)$ and 
the Legendre duplication formula~\cite{ARFKEN2013599} which yields $ \sqrt{\pi }\, 2^{-2p}\,\Gamma(2p+1)  = \Gamma(p+1)  \Gamma \left(p+{\frac {1}{2}}\right)  ,~\forall p \in \mathbb{N}$
and thus
\begin{align}
C_\nu(0) =&= \frac{ 1 }{n\tau }  \sum_{p=0}^{n-1}  \frac{ \Gamma(2p)   }{2^{2p-1}\,\Gamma(p+1)\Gamma(p)} 
= \frac{ 1 }{n\tau \sqrt{\pi}}  \sum_{p=0}^{n-1}  \frac{  \Gamma \left(p+{\frac {1}{2}}\right)  }{\Gamma(p+1)} .
\end{align}%
Using $\Gamma(1/2)=\sqrt{\pi}$, and the monotonicity of the Gamma-function, we find
\begin{align}
( \tau \sqrt{\pi})\,C_\nu(0) &=  ~ 
%
\frac{3\sqrt{\pi}}{n} + \frac{1}{n} \sum_{p=2}^{n-1}  \frac{  \Gamma \left(p+{\frac {1}{2}}\right)  }{\Gamma(p+1)}  
<~ \frac{3\sqrt{\pi}}{n} +\frac{1}{n} \sum_{p=2}^{n-1}  \frac{  \Gamma(p+1)  }{\Gamma(p+1)}
\nn
& =  \frac{3\sqrt{\pi}}{n} -\frac{n-2}{n}  =~ \frac{3\sqrt{\pi}+2}{n} -1.
    	\end{align}
Bessere Grenze finden! Laut Mathematica geht C mit  $C_\nu(0)\sim 1/\sqrt{n}$ gegen null.
\begin{align} 
\displaystyle \lim_{n\rightarrow \infty}\,C_\nu(0)  = 0. 
\end{align}
Second,
}
To this end, we calculate the weight $\Psi= \int_{0}^{\infty} \frac{C_\nu(\Delta t)}{\gamma'  k_\mathrm{B}\mathcal{T}'   k^2}  \mathrm{d}(\Delta t)$of $ C_\nu$, given by 
\begin{align}
\Psi=&  \frac{\tau}{n}  \sum_{p=0}^{n-1}  \sum_{l=0}^{p}\frac{  n^l \tau^{1 }}{2^{2p-l}p!}  \frac{(2p-l)!}{(p-l)!\, l!} 
\, \int_{0}^{\infty}e^{-n{\Delta t}/\tau}\Delta t^l\,\mathrm{d}(\Delta t) 
\nonumber 
\\
= &  \frac{\tau^2}{n^2} \sum_{p=0}^{n-1}  \sum_{l=0}^{p}  \frac{(2p-l)!}{(p!)(p-l)!} \left[\frac{1}{2} \right]^{2p-l}.
\end{align}
Now we use the binomial theorem to further simplify this expression, and find
\begin{align}
\Psi& = \frac{\tau^2}{n^2}\sum_{m=1}^{n}   \sum_{r=0}^{p}  {\binom {p}{r}}  \left[\frac{1}{2}\right]^{r}  \left[\frac{1}{2}\right]^{p-r}
 \nn& =  \frac{\tau^2}{n^2} \sum_{m=1}^{n}   \, \left[\frac{1}{2}+\frac{1}{2}\right]^{2(n-m)} 
 = \frac{\tau^2}{n^2} \sum_{m=1}^{n}  1
=   \frac{\tau^2}{n} .
\end{align}
Hence, the weight \textit{vanishes} as $n\to \infty$. Since $C_\nu$ is obviously a non-negative function of $\Delta t$, this readily implies 
\begin{align}
\lim_{n\rightarrow \infty} C_\nu(\Delta t) =  0.
\end{align}

%
\section{\uppercase{Limit of the heat flow}}\label{sec:Q0limitn}
The limit of the heat flow for $n\to\infty$ corresponds to the heat flow in the system with discrete delay and white noise, which reads~\cite{Loos2019}
\begin{align}
 \dot{Q}_\mathrm{0} =& k^2 \langle  X_n^2\rangle + k a_{00} \langle  X_0(t) X_{0}(t-\tau) \rangle 
 =-a_{00}+k\delta_{\tau,0}
 \nn  
&
-\frac{1 - k \sinh[\frac{\tau}{\gamma}\sqrt{a_{00}^2-k^2}]/\sqrt{|a_{00}^2-k^2|}}{(a_{00} - k\cosh[\frac{\tau}{\gamma}\sqrt{a_{00}^2-k^2}])/(a_{00}^2-k^2)}
.\label{eq:Q0limitn}
\end{align}
%
%
%
%
%
%

\bibliography{AA_references.bib}

\begin{thebibliography}{113}
\bibitem [{not()}]{note3}%
  \BibitemOpen
  \bibinfo {note} {{We have added a nonlinear force $f_0$ in the dynamical Eq.
  of $X_0$~(1), and performed {B}rownian Dynamics simulations to calculate the
  EP. For all cases, we find quadratic increase with $n$.}}\BibitemShut {Stop}%
\end{thebibliography}
\end{document}